\documentclass[sigconf,anonymous=false, 10pt]{acmart}

\usepackage{graphicx}
\usepackage{balance}
\usepackage[utf8]{inputenc}
\usepackage[T1]{fontenc}
\usepackage{bm}
\usepackage{soul}

\usepackage{amsmath}

\usepackage{enumitem}

\usepackage{xcolor} 

\DeclareMathOperator*{\argmin}{arg\,min}

\usepackage{array}
\usepackage{verbatim}

\usepackage{tabularx}

\usepackage{cleveref}
\usepackage{ifthen}
\usepackage{xargs}    

\usepackage[caption=false,font=footnotesize,subrefformat=parens]{subfig}
\usepackage{xspace}
\def\ie{{\textit{i.e.}\xspace}}
\def\eg{{\textit{e.g.}\xspace}}

\def\dst{\textit{destination}\xspace}
\def\env{\textit{environment}\xspace}

\def\sysname{{\sc RadioMic}\xspace}
\def\soundmetric{sound metric\xspace}
\def\dlmodule{RANet\xspace}
\newcommand{\head}[1]{{\noindent \bf #1}}

\renewcommand\footnotetextcopyrightpermission[1]{} 
\setcopyright{none}

\acmDOI{}

\acmISBN{}

\acmConference[]{}
\acmYear{}
\acmPrice{}

\begin{document}

\title{\sysname: Sound Sensing via mmWave Signals}

\author{Muhammed Zahid Ozturk}
\email{ozturk@umd.edu}
\affiliation{%
  \institution{University of Maryland}
  \city{College Park}
  \country{USA}
}
\affiliation{
	\institution{Origin Wireless Inc.}
	\city{Greenbelt}
	\country{USA}
}

\author{Chenshu Wu}
\email{cswu@umd.edu}
\affiliation{%
	\institution{University of Maryland}
	\city{College Park}
	\country{USA}
}
\affiliation{
	\institution{Origin Wireless Inc.}
	\city{Greenbelt}
	\country{USA}
}

\author{Beibei Wang}
\email{bebewang@umd.edu}
\affiliation{%
	\institution{University of Maryland}
	\city{College Park}
	\country{USA}
}
\affiliation{
	\institution{Origin Wireless Inc.}
	\city{Greenbelt}
	\country{USA}
}
\author{K. J. Ray Liu}
\email{kjrliu@umd.edu}
\affiliation{%
	\institution{University of Maryland}%
	\city{College Park}%
	\country{USA}
}
\affiliation{
	\institution{Origin Wireless Inc.}
	\city{Greenbelt}
	\country{USA}
}

\begin{abstract}
Voice interfaces has become an integral part of our lives, with the 
proliferation of smart devices. Today, IoT
devices mainly rely on microphones to sense sound.  
Microphones, however, have fundamental
limitations, such as weak source separation, limited
range in the presence of acoustic insulation, and being prone to
multiple side-channel attacks. 
In this paper, we propose \sysname, a radio-based sound sensing 
system to mitigate these issues and enrich sound applications. 
\sysname constructs sound based on
tiny vibrations on active sources (\eg, a speaker or human throat) or object surfaces (\eg, paper bag), and can work through walls, even a soundproof one. 
To convert the extremely weak sound vibration in the radio signals into sound signals, 
\sysname introduces \textit{radio acoustics}, and presents training-free approaches for robust sound detection and high-fidelity sound recovery. 
It then exploits a neural network to further enhance the recovered sound by expanding the recoverable frequencies and reducing the noises. \sysname translates massive online audios to synthesized data to train the network, and thus minimizes the need of RF data. 
We thoroughly evaluate \sysname under different
scenarios using a commodity mmWave radar. 
The results show \sysname outperforms the state-of-the-art systems significantly. 
We believe \sysname provides new horizons for sound sensing and inspires attractive sensing capabilities of mmWave sensing devices. 
\end{abstract}

\maketitle

\section{Introduction}

\begin{figure}[t]
	\includegraphics[width=0.9\columnwidth]{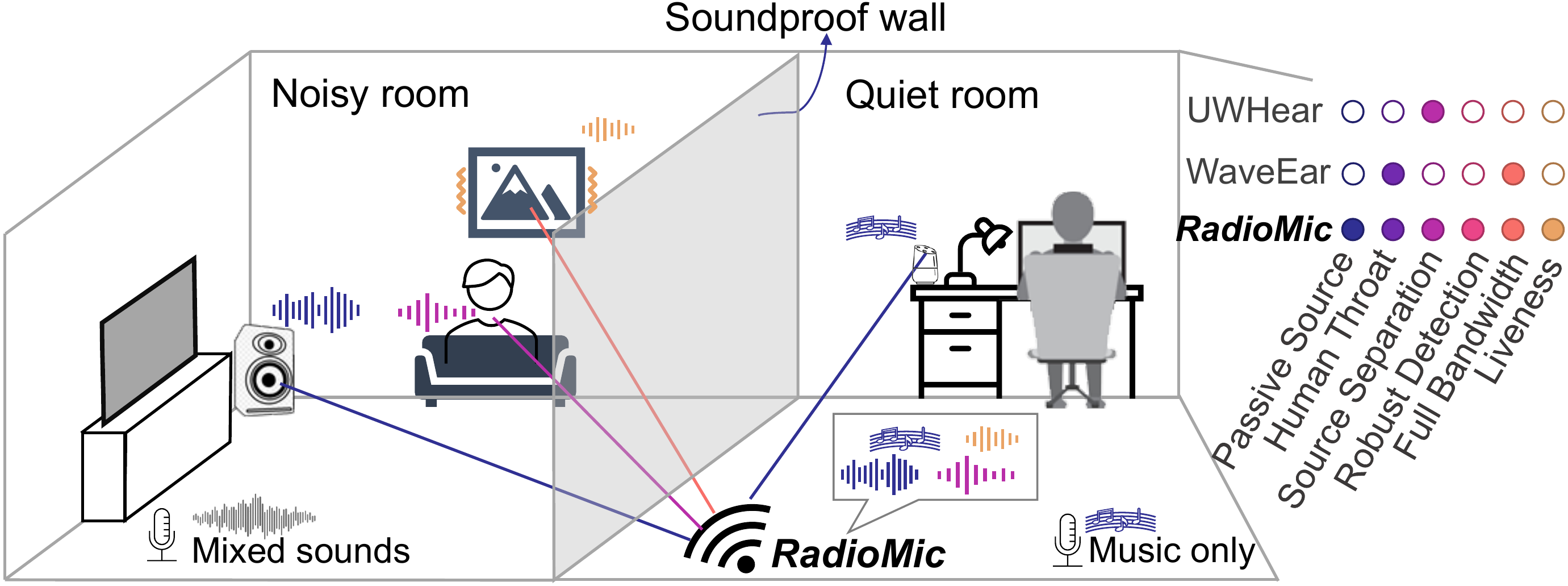}
	\centering
	\caption{An illustrative scenario of \sysname}
	\label{fig:illustration}
\end{figure}

Sound sensing, as the most natural way of human communication, has also become a ubiquitous modality for human-machine-environment interactions. 
Many applications have emerged in the Internet of Things (IoT), 
including voice interfaces, sound events monitoring in smart homes and buildings, acoustic sensing for gestures and health, etc. 
For example, smart speakers like Amazon Alexa or Google Home can now understand user voices, control IoT devices, monitor the environment, and sense sounds of interest such as
glass breaking or smoke detectors, all currently using microphones as the primary interface. 
With the proliferation of such devices in smart environments, 
new methods to sense acoustic contexts have become even more important, while microphones only mimic human
perception, with limited capabilities. 

Various sound-enabled applications acutely demand a next-generation sound sensing modality with more advanced features: 
robust sound separation, noise-resistant, through-the-wall recovery, sound liveness detection against side attacks, etc. 
For example, as illustrated in Fig. \ref{fig:illustration}, robust sound separation can enable a smart
voice assistant to have sustained performance over noisy environments \cite{wang2020uwhear,xu2019waveear}. Being able
to identify animate subjects accurately and quickly would improve the security
of voice control systems against demonstrated audio attacks
\cite{roy2017backdoor, zhang2017dolphinattack,alegre2014re,wu2014signal}. 
Sensing behind an
insulation can increase the operational range of a smart device to multiple
rooms, and allows to retain sound-awareness of outside environments even in a soundproof space. 

Although microphones have been the most commonly used sensors to sense
acoustics events, they have certain limitations. As they can only sense the
sound at the \textit{destination}\footnote{There is a form of microphone, named contact microphone or piezo microphone, that senses sound at the source through contact with solid objects.
} (\ie, the microphone's location), a single microphone cannot separate and identify multiple
sound sources, whereas a microphone array can only separate in the azimuth
direction and require large apertures.
And by sensing any sound arrived at the destination, they raise potential privacy concerns when deploying as a ubiquitous and continuous sound sensing interface in homes. 
In addition, they are prone to inaudible voice attacks and replay attacks, as they only sense the received sound but nothing about the source. 
To overcome some of these limitations, various modalities have been exploited to sense sound signals, like accelerometer \cite{zhang2015accelword}, vibration
motor \cite{roy2016listening}, cameras and light
\cite{davis2014visual, nassi2020lamphone}, laser \cite{muscatell1984laser}, lidar
\cite{sami2020spying}, etc. 
These systems either still sense the sound at the \dst, thus having similar drawbacks as
microphones, or require line-of-sight (LOS) and lighting conditions to operate. 
Consequently, these modalities fail to enable the
aforementioned goals in a holistic way.

Recently, as mmWave has been deployed as standalone low-cost sensing
devices \cite{ti1443mmwave, decawavedw1000},
or on commodity
routers \cite{netgear60ghz}, smartphones \cite{soliphone} and smart hubs
\cite{soli2021sleep}, 
researchers have attempted to sense sound from the \textit{source}
directly using mmWave \cite{wang2020uwhear,xu2019waveear,wei2015acoustic}.
For example, WaveEar \cite{xu2019waveear} employs deep learning with extensive training to reconstruct sound from human throat, using a customized radar. 

UWHear \cite{wang2020uwhear} uses an ultra-wideband radar to extract and separate sound vibration on speaker surfaces. 
mmVib \cite{jiang2020mmvib} monitors single tone machinery vibration using mmWave radar, but not the much weaker sound vibration. 
However, these systems cannot robustly reject non-acoustic motion interferences and mainly reconstructs low frequency sounds ($<$1 kHz), and none of them recovers sound from daily objects like a paper bag. 
Nevertheless, they show the feasibility of sound sensing using mmWave signals and inspire a more advanced design. 

In order to enable the aforementioned features holistically, we propose \sysname, a mmWave-based sensing system that can capture sound and beyond, as illustrated in Figure
\ref{fig:illustration}. 
\sysname can detect, recover and classify sound from sources in multiple
environments.
It can recover various types of sounds, such as music, speech, and environmental sound, from both \textit{active sources} (\eg, speakers or human throats) and \textit{passive sources} (\eg, daily objects like a paper bag). 
When multiple sources present, it can reconstruct the sounds separately 
with respect to \textit{distance}, 
which could not be achieved by classical beamforming in microphone arrays, while being immune to motion interference.
\sysname can also sense sound through walls and even soundproof materials 
as RF signals have different propagation characteristics than sound. 
Potentially, \sysname, located in an insulated room (or in a room with active
noise cancellation \cite{roy2017backdoor}), can be used to monitor and
detect acoustic events \textit{outside} the room, 
offering both sound proofness and awareness in the same time. 
Besides, \sysname can even detect liveliness of a recorded speech and tell whether it is from a human subject or inanimate sources, providing an extra layer of security for IoT devices.

\sysname's design involves multiple challenges. 
1) Sound-induced vibration is extremely weak, on the orders of $\mu$m (\eg, $<10\mu$m on aluminum foil for source sound at $\sim$85dB \cite{davis2014visual}). 
Human ears or microphone diaphragms have
sophisticated structures to maximize this micro vibration. 
Speaker diaphragms or daily objects, however, 
alter the sound vibration differently, and create severe noise, combined with noise from radio devices. 
2) An arbitrary motion in the environment
interferes with the sound signal, and makes robust detection of sound non-trivial, especially when the sound-induced vibration is weak.
3) Wireless signals are prone to
multipath, and returned signals comprise of static reflections, sound vibration,
and/or its multiple copies. 
4) As we will show later, due to the material properties, sounds captured from daily objects are fully attenuated on high frequencies beyond 2 kHz, which significantly impairs the intelligibility of the sensed sounds. 

\sysname overcomes these challenges in multiple distinct ways. 
It first introduces a novel \textit{radio acoustics model} that relates radio signals and acoustic signals. 
On this basis, it detects sound with a training-free module that utilizes fundamental
differences between a sound and any other motion. 
To reduce the effect of background, \sysname 
filters and projects the signal in the complex plane, which
reduces noise while preserving the signal content, 
and further benefits from multipath and receiver
diversities to boost the recovered sound quality.  
\sysname also employs a \textit{radio acoustics neural network} to solve the extremely ill-posed high-frequency reconstruction problem, 
which leverages massive online audio datasets and requires minimal RF data for training. 

We implement \sysname using a commercial off-the-shelf (COTS) mmWave
radar, evaluate the performance and compare with the state of the arts
 in different environments, using diverse sound files
at varying sound levels. 
The results show that \sysname can recover sounds from active sources such as speakers and human throats, and passive objects like aluminum foil, paper bag, or bag of chips. 
Particularly, \sysname outperforms latest approaches \cite{wang2020uwhear,jiang2020mmvib} in sound detection and reconstruction under various criteria. 
Furthermore, we demonstrate case studies of multiple source separation and sound liveness detection to show interesting applications that could be achieved by \sysname. 

In summary, our contributions are:
\begin{itemize}[leftmargin=1em]
	\item We design \sysname, an RF-based sound sensing system that separates multiple sounds and operates through the walls. To the best of our knowledge, \sysname is the first RF system that can recover sound from passive objects and also detect liveness of the source. 
	\item We build the first radio acoustics model from the perspective of Channel Impulse Response, which is generic to the underlying RF signals and underpins training-free robust sound detection and high-fidelity sound reconstruction. 
	\item We develop a radio acoustics neural network, requiring minimal RF training data, to enhance the sensed sound by expanding the recoverable frequencies and denoising. 
	\item We implement \sysname on low-cost COTS hardware and demonstrate multiple attractive applications. The results show that it outperforms the state-of-the-art approaches. 

\end{itemize}

\section{Radio Acoustics}
\label[]{sec:dopplervibration}
In this section, we explain mechanics of the sound sensing using radio
signals, which we name as \textit{radio
acoustics}. 

\begin{figure}[t]
	\includegraphics[width=0.38\textwidth]{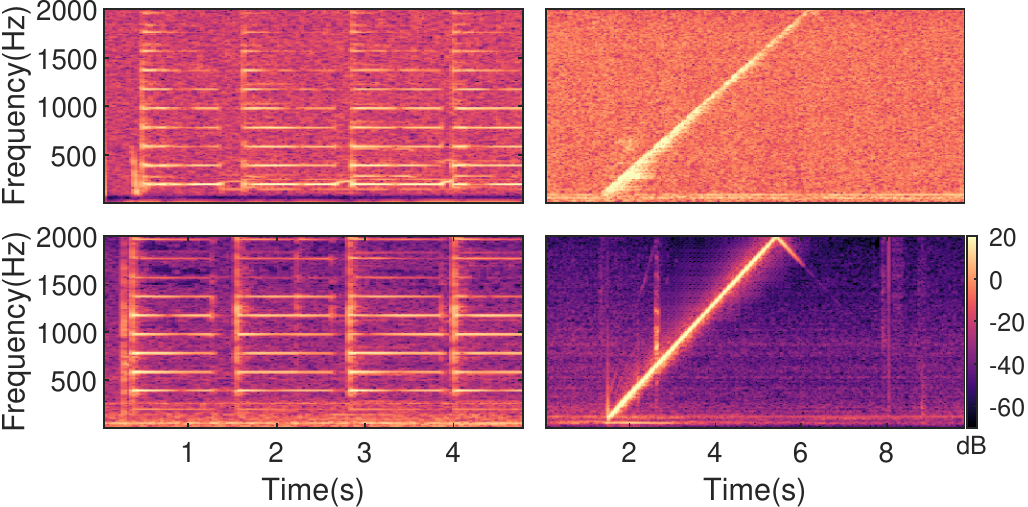}
	\centering
	\caption{Examples of radio-sensed sound. \rm{Bottom are microphone references. Left: guitar sound (Note G3, 196 Hz); Right: frequency sweep sound sensed from aluminum foil.}}
	\label{fig:guitarspectrum}
\end{figure}

\subsection{Mechanics}
Sound is basically modulation of medium\footnote{Without loss of generality, we assume the medium to be air in this paper.} pressure through various
mechanisms. 
It is generated by a vibrating surface, and the modulation signal travels through
in place motion of air molecules. 

A vibrating surface could be a speaker diaphragm, human throat, strings of
musical instrument such as a guitar, and many daily objects like a paper bag. 
In the case of speakers, motion on the speaker diaphragm modulates the signal,
whereas in human throat, vocal cords create the vibration, with the mouth and
lips operating as additional filters, based on the source-filter model
\cite{fant1970acoustic}. 
To sense the sound, the same mechanism is employed at the microphones to convert the changes in the air pressure into
electrical signal, via suitable diaphragms and electrical circuitry. 
Microphone diaphragms are designed to be sensitive to air vibration and optimized to capture the
range of audible frequencies (20Hz-2kHz), and even beyond \cite{roy2017backdoor}. 

Mechanics of extracting sound from radio signals rely on the Doppler phenomenon and the relationship between the object vibration and the reflected signals. 
The vibration alters how the signals reflected off the object surface propagate. 
Therefore, the reflection signals can measure the tiny vibrations of an object surface in terms of Doppler shifts, from which we can recover the sound.

The vibration not only occurs at the source where sound is generated, but also on intermediary objects that are incited by the air. 
Most commonly, sound-modulated air molecules can cause $\mu$m
level or smaller vibration on any object surface. 
The vibration amplitude depend on
 material properties and various factors. 
Generally, sound vibrations are stronger at the source where they are generated (referred as \textit{active vibration
sources}), and much weaker at the intermediary objects (referred as \textit{passive vibration sources}). 
Microphones typically only sense sound from active sources, as the air-modulated passive sound is too weak to further propagate to the microphone diaphragms. 
Differently, as radio acoustics sense sound directly at the source, it can potentially reconstruct sound from both active sources and passive sources. 
To illustrate the concept, we place a radar in front of a guitar, and
play the string G3 repeatedly, and provide the radio and microphone
spectrograms in Fig. \ref{fig:guitarspectrum}. As shown, when the string is hit, it continues to vibrate (or move back
and forth) in place to create the changes in the air pressure, and therefore the
sound. The radar senses the sound by capturing the motion
of the strings, whereas the microphone captures the modulated air pressure at
the 
diaphragm. 
The right part of Fig. \ref{fig:guitarspectrum} shows an example of sound sensed from an aluminum foil. 
Although the sensing mechanisms of microphones and radio
are completely 
different in their nature, they capture the same phenomenon, and the resulting
signals are similar. 

\subsection{Theoretical Background}
\label[]{sec:theory}
Starting with the mechanics of sound vibration, we build the radio acoustics model. Later on, we explain how this model
is used to reconstruct sound from radio signals in \S\ref{sec:systemdesign}.

\head{Sound Vibration Properties:} 
As explained previously, sound creates vibration (motion) on objects, which is proportional to the transmitted energy of sound
from the air to the object and depends on multiple factors, such as inertia
and signal frequency \cite{fahy2000foundations}. Denoting the acoustic signal
with $a(t)$, we can model the displacement due to sound as:
\begin{equation}
	x(t) = h\star a (t),
	\label[]{eq:freqresp} 
\end{equation}
where $h$ denotes vibration generation mechanism for an active source or the impulse response of the air-to-object interface for a passive object, 
and $\star$ represents
convolution. 

\head{Radio Acoustic Model:} 
\label[]{subsec:signalmodel}
From the point view of physics, sound-induced vibration is identical to machinery vibration, except that sound vibration is generally of magnitude weaker. 
To model the sound vibration from RF signals, one could follow the model used in \cite{jiang2020mmvib}, which however assumes the knowledge of the signal model and thus depends on the specific radio devices being used. 
In \sysname, we establish a model based on the Channel Impulse Response (CIR) of RF signals, which is generic and independent of the underlying signals. 
By doing so, in principle, the model applies to any radio device that outputs high-resolution CIR, be it an FMCW radar \cite{jiang2020mmvib}, an impulse radar \cite{wang2020uwhear}, or others \cite{palacios2018adaptive}.

\begin{figure*}[t]
	\includegraphics[width=0.85\textwidth]{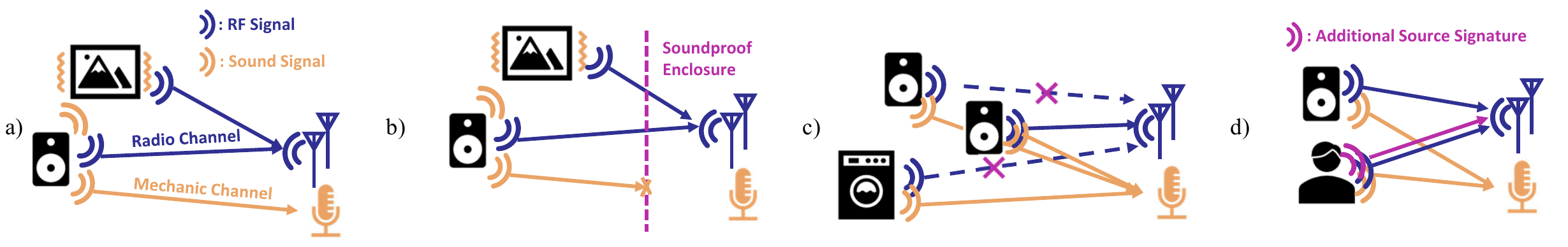}
	\centering
	\caption{Different use cases of \sysname. \rm{a) Sensing sound from active/passive sources, b) Sensing through soundproof materials, 
	c) Separating sound of multiple sources, d) Sound liveness detection}}
	\label{fig:systemusescenarios}
\end{figure*}

CIR of an RF signal can be given as 
\begin{equation}
	g(t,\tau) = \sum_{l=0}^{L-1} \alpha_{l}{(t)}\delta(\tau - \tau_l(t)),
	\label[]{equ:cirmodel}
\end{equation}
where $t$ and $\tau$ are referred as long and short time, $L$ denotes number of range bins (sampling wrt. distance), $\alpha_l$ denotes complex scaling factor, $\tau_l$ is the roundtrip duration from range bin $l$, and $\delta(\cdot)$ represents dirac delta function, indicating presence of an object. Assuming no multipath, and an object of interest at range bin $l^*$, corresponding to time delay $\tau^{*}$, CIR of that range bin can be given as:
\begin{equation}
    g(t,\tau^*) = \alpha_{l^{*}}{(t)} \exp(-j2\pi f_c\tau_l^{*}(t)),
    \label[]{equ:vibrationmodel}
\end{equation}
where $f_c$ denotes the carrier frequency. If we assume the object to remain
stationary in range bin $l^{*}$, we can drop the variables $\tau^*$, and
$l^{*}$, convert time delay into range, and rewrite the CIR as:
\begin{equation}
    g(t) = \alpha{(t)} \exp(-j2\pi R(t)/\lambda)),
    \label[]{equ:vibrationmodelsimple}
\end{equation}
where $R(t)$ denotes the actual distance of the object, and $\lambda$ denotes the wavelength.

Now, considering a vibrating object (\ie, sound source), we can decompose the range value into the static and vibrating part as
$R(t) = R_0 + x(t)$. 
As can be seen, there is a direct relationship between the CIR $g(t)$ and the phase of the returned
signal. By extracting the phase, $g(t)$ could be used to derive $R(t)$, and therefore, the vibration
signal, $x(t)$. We further omit the temporal dependency of $\alpha$, as we assume the object to be stationary, and the effect of displacement due to vibration on path loss to be negligible.

So far, we have assumed to have the vibrating object in the
line of the radar solely, and did not account for other reflections from the environment.
As suggested by \eqref{equ:vibrationmodelsimple}, $g(t)$ lies on a circle in the $IQ$ plane with center at the origin. However,
due to various background reflections, $g(t)$ is actually superimposed with a
background vector, and the circle center is shifted from the origin. Thus, $g(t)$ can be written as:
\begin{equation}
	g(t) = \alpha\exp\left(-j2\pi \tfrac{R(t)}{\lambda}\right) + \alpha_B(t)\exp(j\gamma(t)) + w(t),
	\label[]{eq:completesignal}
\end{equation}
where $\alpha_B(t)$ and $\gamma(t)$ are the amplitude and phase shift caused
by the sum of all background reflections and vibrations, and $w(t)$ is the
additive white noise term. Equation \eqref{eq:completesignal} explains the received signal model, and will be used to build our reconstruction block of \sysname in \S\ref{sec:systemdesign}.

\subsection{Potential Applications}
\label[]{subsec:applicationscenarios}

As in Fig. \ref{fig:systemusescenarios}, \sysname could benefit various applications, including many that have not been easily achieved before. 
By overcoming limitations of today's microphone, \sysname can enhance performance of popular smart speakers in noisy environments. 
Collecting spatially-separated audio helps to better understand acoustic events of human activities, appliance functions, machine states, etc. 
Sound sensing through soundproof materials will provide awareness of outside contexts while preserving the quiet space, which would be useful, for example, to monitor kid activities while working from home in a closed room. 
Detect liveness of a sound source can protect voice control systems from being attacked by inaudible voice \cite{zhang2017dolphinattack} or replayed audio \cite{wu2014signal}. 
With mmWave entering more smart devices, \sysname could also be combined with a microphone to leverage mutual advantages. 

On the other hand, \sysname can be integrated with other existing wireless sensing applications. 
For example, Soli-based sleep monitoring \cite{soli2021sleep} currently employs microphone to detect coughs and snore, which may pose privacy concerns yet is no longer needed with \sysname. 
While remarkable progress has been achieved in RF-based imaging \cite{zhao2018rf,jiang2020towards,zhang2020mmeye}, \sysname could offer a channel of the accompanying audio. 

We will show proof of concepts for some of the applications (\S\ref{sec:casestudies}) and leave many more to be explored in the future.

\section{RadioMic Design}
\label[]{sec:systemdesign}

\sysname consists of four main
modules. 
It first extracts CIR from raw RF signals (\S\ref{subsec:rawsignal}). From there, it detects sound vibration (\S\ref{sec:sounddetectionlocalization}) and recovers the sound (\S\ref{sec:soundspectrumextraction}), while rejecting non-interest motion. 
Lastly, it feeds the recovered sound into a neural network for enhancement (\S\ref{subsec:neuralnetwork}). 

\subsection{Raw Signal Conversion}
\label[]{subsec:rawsignal}

Our implementation mainly uses a COTS FMCW mmWave radar, although \sysname can work with other mmWave radios that report high-resolution CIR such as an impulse radar. 
Prior to explaining how \sysname recovers sound, we provide preliminaries to extract CIR from a linear
FMCW radar for comprehensive explanation. 

CIR on impulse radar has also been exploited \cite{wang2020uwhear,zhang2020mmeye,wu2020msense}. 

An FMCW radar transmits a single tone signal with linearly
increasing frequency, called a chirp, and captures the echoes from the environment.
Time delay of the echoes could be extracted by
calculating the amount of frequency shift between the transmitted and received signals, which can be converted to a range information.
This range information is used to differentiate an object from the other reflectors in the environment.
In order to obtain the range information, the frequency shift between
transmitted and received signals are calculated by applying FFT, which is
usually known as \textit{Range-FFT} \cite{stove1992linear}. The output of
Range-FFT can be considered as CIR, $g(t,\tau)$, and our modeling in
\S\ref{sec:theory} would be applicable. 

\sysname further obtains the so-called \textit{range-Doppler spectrogram} from the CIR, 
which can be extracted by a short-time Fourier Transform (STFT) operation. 
STFT is basically FFT operations applied in the $t$ dimension in
$g(t,\tau)$ for subsets of long-time indices, called \textit{frames}.
We denote the output range-Doppler spectrograms as $G(f,r,k)$, where $f \in (-N_s/2,N_s/2)$ denotes frequency shift, $r$ corresponds to range bins (equivalent to $\tau_l$), and $k$ is the frame index. 
We note that $G$ is defined for both positive and negative frequencies, corresponding to different motion directions of the objects, which will be used in the following section.

\subsection{Sound Detection \& Localization}
\label[]{sec:sounddetectionlocalization}
As any range bin can have sound vibration, it is critical to have a
robust detection module that can label both \textit{range bins} and \textit{time
indices} effectively. 
Standard methods in the literature, such as constant
false alarm rate (CFAR)
or Herfindahl–Hirschman
Index (HHI) \cite{wang2020uwhear} are not robust (See \S\ref{subsec:evaldetectionlocalization}), as we envision a system to be triggered \textit{only} by sound vibration but not arbitrary motion. 

In \sysname, we leverage the physical properties of sound vibration to design a new approach. 
Mainly, \sysname relies on the fact that a
vibration signal creates both \textit{positive} and \textit{negative} Doppler
shifts, as it entails consequent displacement in both directions. This forward and backward motion is expected
to have the same amplitudes at the same frequency bins, but with the opposite signs as the total displacement is zero. This would result in
\textit{symmetric spectrograms}, as also noted by other work \cite{rong2019uwbradar,jiang2020mmvib}. 
\sysname exploits this observation with a novel metric for robust sound detection. 

To define a sound metric, let
$G^+(f,r,k)$ denote the magnitude of the positive frequencies of range-Doppler spectrogram $G(f,r,k)$, \ie, $G^+(f,r,k) = |G(f,r,k)| \text{ for } f\in(0,N_s/2)$. 
Similarly, we define
$G^-(f,r,k)$ for negative frequencies. Note that the values of $G^+$ and $G^-$ are always positive, as they are defined as
magnitudes, and would have non-zero mean even when there is no signal, due to the additive noise. 
Calculating the cosine distance or correlation coefficient would result in high
values, even if there is only background reflections. In order to provide a more robust metric, we subtract the noise floor from both $G^+$ and $G^-$ and denote the resulting matrices with $\hat{G}^+$ and $\hat{G}^-$.
Then, instead of using standard cosine distance, we change the definition to
enforce similarity of the amplitude in $\hat{G}^+$ and $\hat{G}^-$:  
\begin{equation}
	m(r,k) = \frac{\sum_f |\hat{G}^{+}(f,r,k)\hat{G}^{-}(f,r,k)|^2}{\max{\left(\sum_f 
	|\hat{G}^{+}(f,r,k)|^2,\sum_f |\hat{G}^{-}(f,r,k)|^2\right)}}.
	\label[]{equ:soundmetric}
\end{equation}

 \begin{figure}[t]
 	\includegraphics[width=0.9\columnwidth]{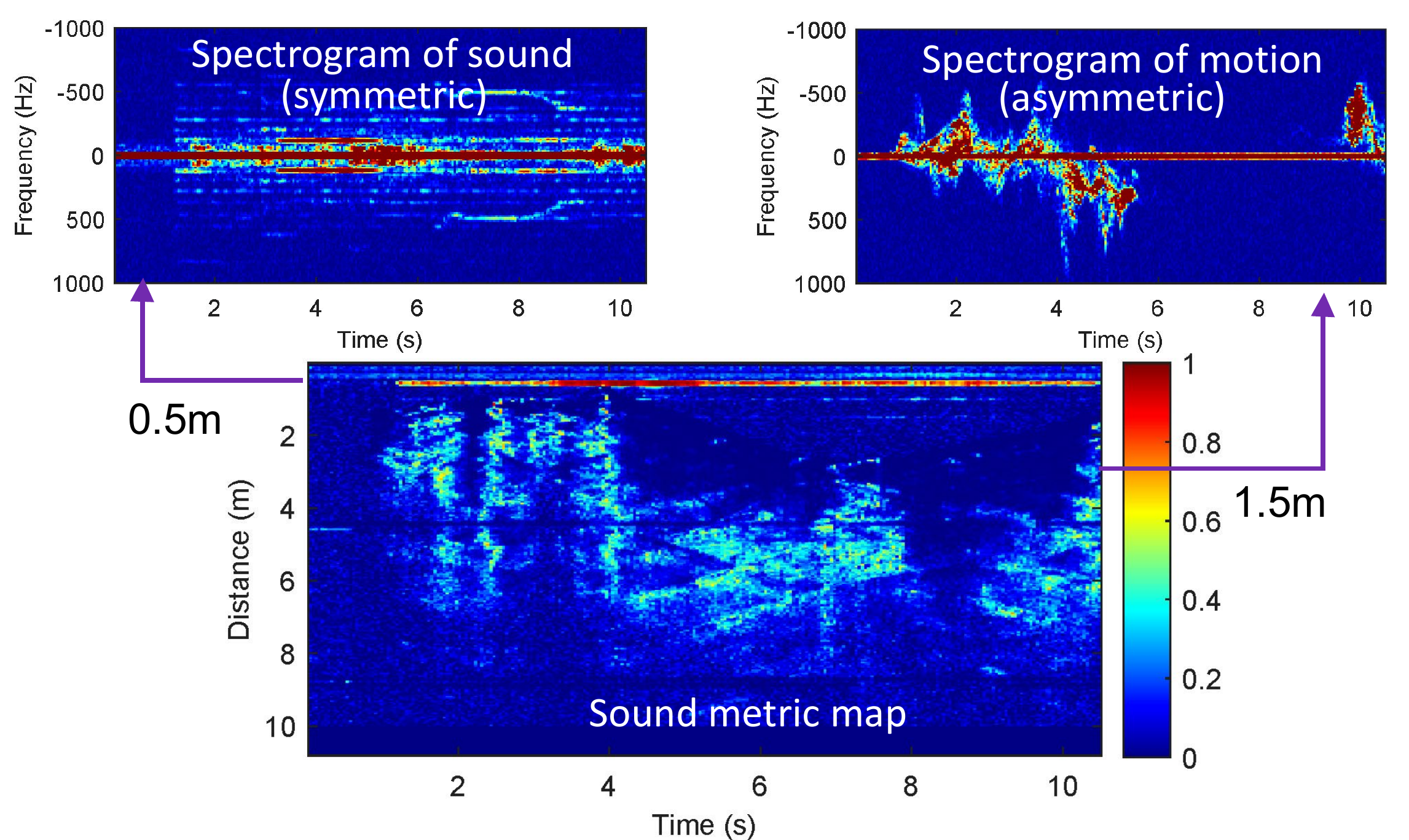}
 	\centering
 	\caption{Sound metric. \rm{An aluminum foil is placed at 0.5m. Music starts playing around 1.5s, while random motion occurs at distances 1$\sim$3m for 10s. \textit{Spectrograms} at distance (a) 0.5m and (b) 1.5m, and (c) resulting \textit{sound metric map}}}
 	\label{fig:dopplersymmetric}
 \end{figure}

\sysname calculates the \textit{sound 
metric} as in \eqref{equ:soundmetric} for each range bin $r$, and for each time-frame $k$, resulting in a \textit{sound metric map} as illustrated in in Fig. 
\ref{fig:dopplersymmetric}c, 
music sound (Fig.\ref{fig:dopplersymmetric}a) results in high values of the sound metric, whereas arbitrary motion (Fig.\ref{fig:dopplersymmetric}b) is
suppressed significantly, due to asymmetry in the Doppler
signature and power mismatches. This illustrates the responsiveness of sound metric to vibration, 
while keeping comparatively lower values for random motion. 

To detect
vibration, \sysname uses a median absolute deviation based outlier detection
algorithm, and only extracts outliers with positive deviation. Our evaluation in
\S\ref{sec:analysis} shows that, an outlier based scheme outperforms a
fixed threshold, as non-sound motion sometimes can arbitrarily have high values, which can create false alarms for a hard threshold.
Additionally, this approach also adapts perfectly to various motion and sounds of diverse amplitudes, including those from active sources and passive sources.

As the sound detection algorithm runs on the range bins, it can detect
multiple range bins with sound. The radio signals in these range bins will
be processed separately by \sysname. This enables detection of multiple
sources and reconstruction of each sound signal separately. 
As a byproduct, it also locates at which bin does the sound occur, and reduces interference. 

\subsection{Sound Recovery}
\label[]{sec:soundspectrumextraction}
Having extracted the time indices and the range information about active or
passive vibration sources in the environment, \sysname extracts raw sound signals. 
Using the signal model in \eqref{eq:completesignal}, \sysname recovers
the acoustic signal by first filtering out the interference and background and
approximating the remaining signal with a line fit to further reduce noise.

We first apply an FIR high-pass filter as the changes in the background usually have much lower frequencies. 

The resulting signal, $\hat{y}(t)$ can be given as:
\begin{equation}
	\hat{g}(t) \approx \alpha\exp\left(-j2\pi \tfrac{R(t)}{\lambda}\right) - \alpha\exp(j\gamma_R) + \hat{w}(t),
\end{equation}
where $\hat{w}(t)$ is the filtered noise term, and $\alpha \exp(j\gamma_R)
\approx \alpha \exp(-j2\pi\tfrac{R_0}{\lambda})$ is the center of a circle. 
The
signal component, $\exp(-jg\pi \tfrac{R(t)}{\lambda})$, remains mostly unchanged, due to the frequencies of interest with
sound signals, and this operation moves the arc of the vibration circle to the
origin in the IQ plane. Furthermore, this operation reduces any drifting in IQ plane, caused by the hardware.

As explained in the previous section, the
curvature of the arc is in the order of $1^\circ$ for $\mu m$ displacement with a mmWave device, by projecting the arc, $\alpha\exp(-j2\pi \tfrac{R(t)}{\lambda})$, onto
the tangent line at $\alpha\exp(-j2\pi \tfrac{R_0}{\lambda})$, we can approximate $\hat{g}(t)$ as
\begin{equation}
	\hat{y}(t) = m + nx(t) + \hat{w}(t) - \alpha\exp(j\gamma_R),
	\label[]{eq:linearfitsimple}
\end{equation}
where $m=\exp(-2\pi\tfrac{R_0}{\lambda})$, and $n=\alpha\tfrac{-2\pi}{\lambda}
\exp(-j(\tfrac{\pi}{2}+2\pi \tfrac{R_0}{\lambda}))$. 

Using the fact that $\alpha_R \exp(j\gamma_R) \approx \alpha
\exp(-j2\pi\tfrac{R_0}{\lambda})$, $\hat{g}(t)$ can be further simplified as
\begin{equation}
	\hat{g}(t) \approx nx(t) + \hat{w}(t),
\end{equation}
which suggests that, $\hat{g}(t)$ already has the real-valued sound signal
$x(t)$, scaled
and
projected in complex plane, with an additional noise. Using the fact that the
projection onto an arbitrary line does not change noise variance, we can
estimate $x(t)$ with minimum mean squared error (MMSE) criteria. The estimate is given as:
\begin{equation}
	\hat{x}(t) = \mathcal{R}\{\hat{g}(t)\exp(-j\hat{\theta})\},
\end{equation}
where $\mathcal{R}$ is the real value operator. Angle
$\hat{\theta}$ can be found as:
\begin{equation}
	\hat{\theta} = \argmin_{\theta} \|\hat{g}(t) - \mathcal{R}\{\hat{g}(t)\exp(-j\theta)\}\exp(j\theta)\|^2.
\end{equation}

In order to mitigate any arbitrary motion and reject noise, we first project
the samples on the optimum line, then calculate spectrogram. Afterwards, we
extract the maximum of two spectrograms (\ie~$\max\{G^+, G^-\}$) and apply inverse-STFT to construct
real-valued sound signal. This is different than the naïve approaches such as
extracting maximum of the two spectrograms immediately \cite{rong2019uwbradar},
or extracting one side only \cite{xu2019waveear} as we first reduce the noise
variance by projection. This
also ensures a valid sound spectrogram, as the inverse spectrogram needs to
result in a \textit{real-valued sound signal}. 

\head{Diversity Combining:} 
The raw sound output does not directly result in very high
quality sound, due to various noise sources and the extremely small
displacement on the object surface. 
To mitigate these issues, we utilize two physical redundancies in the system: 
1) \textit{receiver diversity} offered by multiple antennas that can be found in many radar arrays, and 2) \textit{multipath diversity}.
Multiple receivers sense the same vibration signal from slightly
different locations, and combination of
these signals could enable a higher signal-to-noise ratio (SNR).
In addition, multipath environment enables to observe a similar sound signal in
multiple range bins, which could be used to further reduce the noise levels. To
exploit these multiple diversities, we employ a selection combining scheme.
Mainly, silent moments are used to get an estimate for each multipath component
and receiver. When sound is detected, the signal with the highest SNR is
extracted. 
We employ this scheme, as we sometimes observe significant differences between received signal quality on different antennas, due to hardware noise (\eg some antennas report noisier outputs, occasionally), and specularity effect.
We only consider nearby bins when combining multipath components, in order to recover sound from multiple sources.

\subsection{Sound Enhancement via Deep Learning}
\label[]{subsec:neuralnetwork}
Even though the aforementioned processes reduce multiple noise sources, and optimally create a sound signal from radio signals, the recovered sound signals face two issues: 
\begin{itemize}[leftmargin=1em]
	\item Narrowband: \sysname so far cannot recover frequencies beyond 2 kHz, noted as \textit{high-frequency deficiency}. This creates a
	significant problem, as the articulation index, a measure of the amount of
	intelligible sound, is less than 50\% for 2kHz
	band-limited speech, as reported by \cite{french1947factors}.
	\item Noisy: The recovered sound is a noisy copy of the original sound. As
	the vibration on object surfaces is on the order of $\mu m$, phase noise
	and other additive noises still exist.	
\end{itemize}

\head{High-Frequency Deficiency:} 
As our results show, frequencies beyond 2 kHz are attenuated fully in the recovered sound, as the channel $h$ in \eqref{eq:freqresp} removes
useful information in those bands. 
To explain why this happens, we return back to our modeling of the signal. Namely, the output signal $\hat{x}(t)$ is
a noisy copy of $x(t)$, which could be written as:
\begin{equation}
	\hat{x}(t) \approx x(t) + \hat{w}(t) = h \star a(t) + \hat{w}(t),
	\label[]{eq:recoveryestimate}
\end{equation}
from \eqref{eq:freqresp}. As can be seen, what we observe is the output of the
air pressure-to-object vibration channel (or mechanical response of a speaker),
as also observed by \cite{davis2014visual, roy2016listening, sami2020spying},
In order to recover $a(t)$ fully, one
needs to invert the effect of $h$.
However, classical signal processing techniques like spectral subtraction or equalization cannot recover the entire
band, as the 
information at the high frequencies have been lost, and these classical methods
cannot exploit temporal redundancies in a speech or sound signal.

\head{Neural Bandwidth Expansion and Denoising:} 
To overcome these issues, namely, to reduce the remaining noise and reconstruct the high frequency part, we resort to deep learning. We
build an autoencoder based neural network model, named as \textit{radio acoustics networks} (\dlmodule), which is modified from the classical UNet
\cite{ronneberger2015unet}. Similar models have been proposed
for bandwidth expansion of telephone sound (8kHz) to high fidelity sound (16
kHz) \cite{lagrange2020bandwith,li2015deep,abel2018artificial} and for audio
denoising \cite{park2016fully,xu2014experimental}. Although the formulation of the problem
is similar, theoretical limitations are stricter in \sysname, as
there is severer noise, and stronger band-limit constraints on the recovered speech (expanding 2 kHz to 4 kHz), in addition to the need for solving both problems together. 

\begin{figure}
 	\includegraphics[width=0.9\columnwidth]{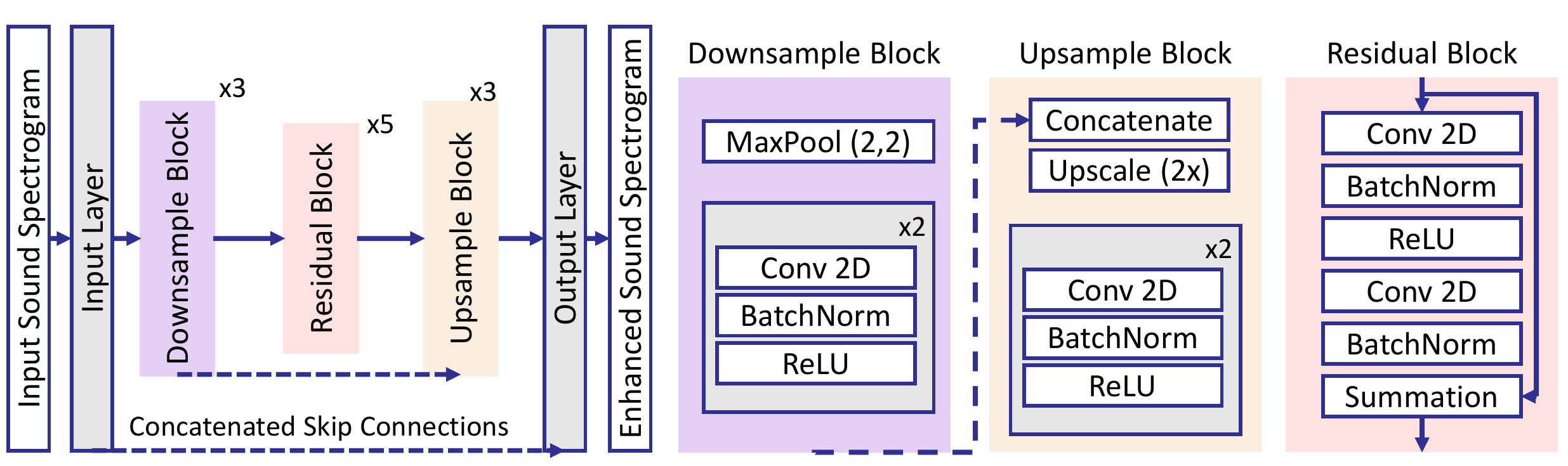}
 	\centering
 	\caption{\dlmodule Structure}
 	\label{fig:neuralnetstructure}
 \end{figure}

\begin{figure}
	\includegraphics[width=0.9\columnwidth]{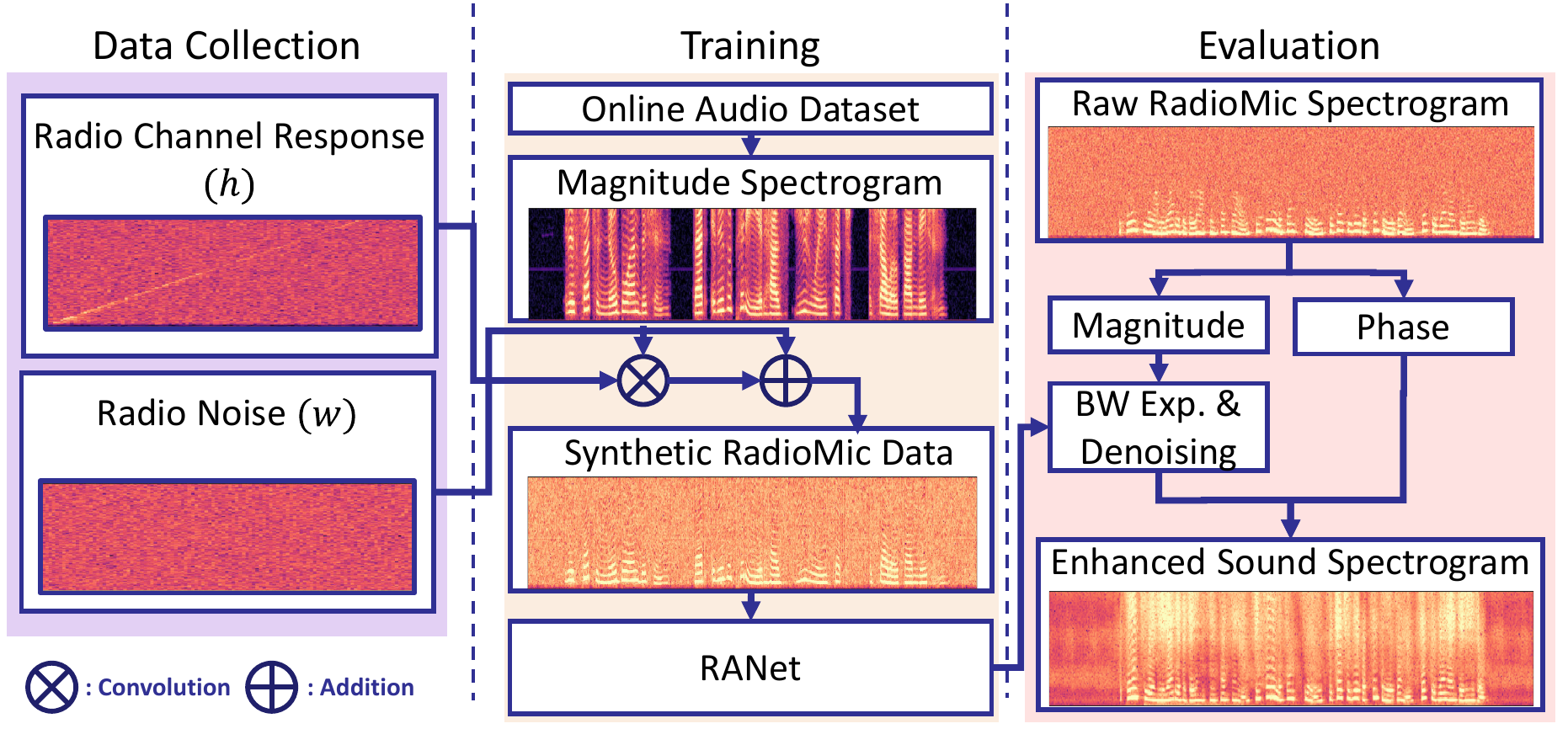}
	\centering
	\caption{Working process of \dlmodule in \sysname}
	\label{fig:neuralnetflow}
\end{figure}

Fig. \ref{fig:neuralnetstructure} portrays the structure of \dlmodule, with the entire processing flow of data augmentation, training, and evaluation illustrated in Fig. \ref{fig:neuralnetflow}. 
We highlight the major design ideas below and leave more implementation details to \S\ref{sec:implementation}.

\textit{1) \dlmodule Structure: }
\dlmodule consists of downsampling, residual and upsampling blocks, which are connected sequentially, along with some residual and skip connections. On a high level, the encoding layers (downsampling blocks) are used to estimate a
latent representation of the input spectrograms (\eg~similar to images); and decoding
layers (upsampling blocks) are expected to reconstruct high-fidelity sound.
Residual layers in the middle are added to capture more temporal and spatial dependencies by increasing the receptive field of the convolutional
layers, and to improve model complexity.
\dlmodule takes input spectrograms of size $(128\times128)$, as
grayscale images, and uses $(3\times3)$ strided convolutions, with number of
kernels doubling in each layer of downsampling blocks, starting from $32$. In the
upsampling blocks, the number of kernels are progressively reduced by half, to
ensure a symmetry between the encoder and decoder. In the residual blocks,
the number 
of kernels do not change, and outputs of each double convolutional layer is
combined with their input. We use residual and skip connections
to build \dlmodule, as these are shown to make the training
procedure easier \cite{he2015deep}, especially for deep neural networks. 

\textit{2) Training \dlmodule without Massive RF Data: }
	As we have proposed a relatively deep
	neural network for an extremely challenging inverse problem, a successful
	training process
	requires extensive data collection. 
	However, collecting massive RF data is costly, which is a practical limitation of many learning-based sensing systems. 
	On the other hand, there have seen a growing, massive audio datasets becoming available online. 
	In \sysname, instead of going through an extensive data collection procedure like \cite{xu2019waveear}, 
	we exploit the proposed radio acoustics model and 
	translate massive open-source datasets to synthetically simulated radio sound for training. 
	Two parameters are particularly needed to imitate radio sound with an audio dataset, 
	\ie, the channel $h$ and noise $w$ as in \eqref{eq:recoveryestimate}.
	We use multiple estimates for these parameters to cover different scenarios, and artificially create radar sound at different noise levels and for various
	frequency responses, thus allowing us to train \dlmodule efficiently with very little
	data 
	collection overhead. Furthermore, this procedure ensures a rigorous system
	evaluation, as only the evaluation dataset consists of the \textit{real}
	radio speech.

\textit{3) Generating Sound from \dlmodule:} 
Using the trained model, \dlmodule uses raw
radar sound as input, extracts magnitude spectrograms that will be used for
denoising and bandwidth expansion. Output magnitude spectrograms of \dlmodule is combined
with the phase of the input spectrograms, as usually done in similar work
\cite{xu2014experimental,boll1979suppression} and the
time-domain waveform of the speech is constructed. In order to reduce the
effect of padding on the edges, and capture long-time dependencies, only the
center parts of the estimated spectrograms are used, and inputs are taken as
overlapping frames with appropriate paddings in two sides, similar to \cite{park2016fully}.

\section{System Implementation}
\label[]{sec:implementation}

\head{Hardware Setup}
\label[]{subsec:hardwaresetting}
While \sysname is not limited to a specific type of radar, we mainly use an FMCW mmWave radar for our implementation.
We use a COTS mmWave radar TI IWR1443 with a real-time data capture board DCA1000EVM, both produced by Texas Instruments. It works on 77GHz with a bandwidth of 3.52GHz. 
The evaluation kit costs around \$500, while the chip could
be purchased for \$15.

Our radar device has 3 transmitter (Tx) and 4 receiver (Rx) antennas. 
We set the sampling rate of 6.25
kHz, which enables to sense up to 3.125 kHz, close to human sound range, while avoiding too high duty cycle on the radar.
We use two Tx antennas, as the device does not allow to utilize all the three at this configuration. 
This enables 8
virtual receivers, due to the placement of antennas. 
We do not use customized multi-chirp signals as in \cite{jiang2020mmvib} since we found it does not benefit much while may prohibit the potential for simultaneous sensing of other applications. 

\head{Data Processing:}  
Using the software provided by Texas Instruments, and applying Range-FFT, extracted CIR results in
($8\times256\times6250$) samples per second. 
We extract range-Doppler spectrograms using frame lengths of 256 samples
($\approx 40$ ms) with $75\%$ overlap, and periodic-Hanning window. 
The selection of
window function ensures perfect reconstruction, and has good sidelobe
suppression properties, to ensure reduced effect of DC sidelobes on symmetry of range-Doppler spectrograms.

\begin{figure*}[t]
 	\centering
 \begin{minipage}{0.23\textwidth}
 	\includegraphics[width=1.0\textwidth]{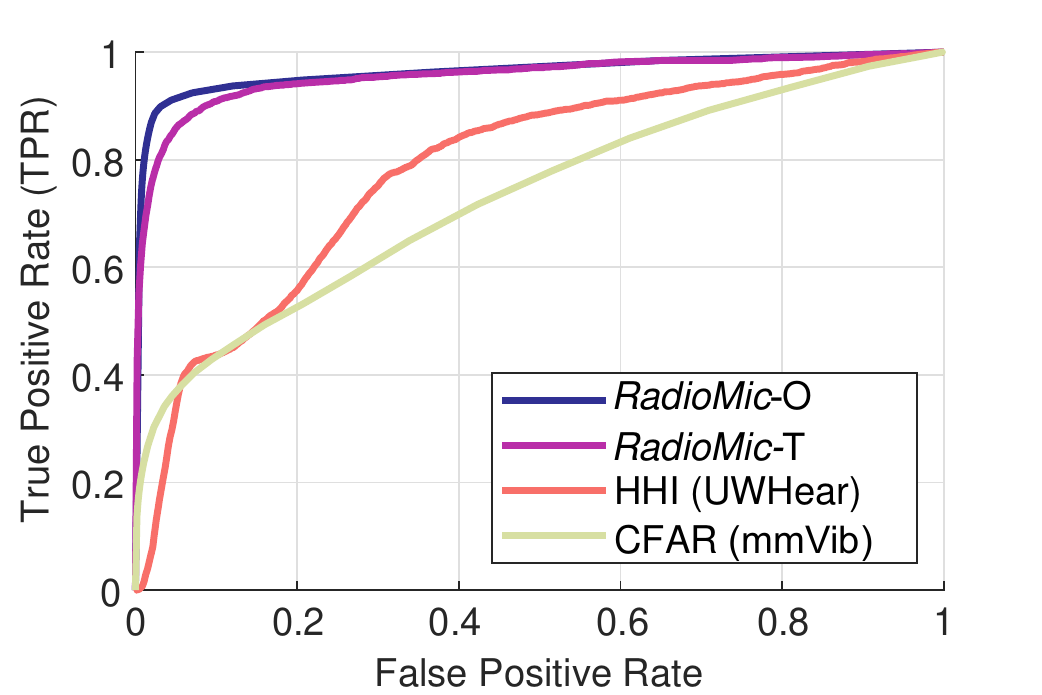}
	\centering
	\caption{ROC curve of sound detection}
	\label{fig:detectorroc}
 \end{minipage}
 \hfill
 \begin{minipage}{0.225\textwidth}
	\includegraphics[width=1.0\textwidth]{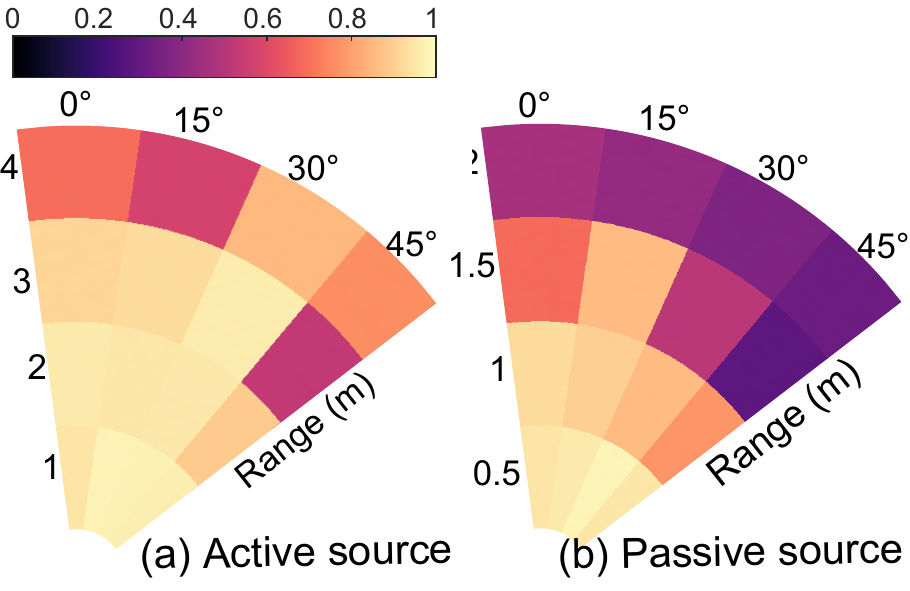}
	\caption{Detection coverage of \sysname}
	\label{fig:detection_heatmap_overall}
 \end{minipage}
 \hfill
 \begin{minipage}{0.23\textwidth}
 	\includegraphics[width=1.0\textwidth]{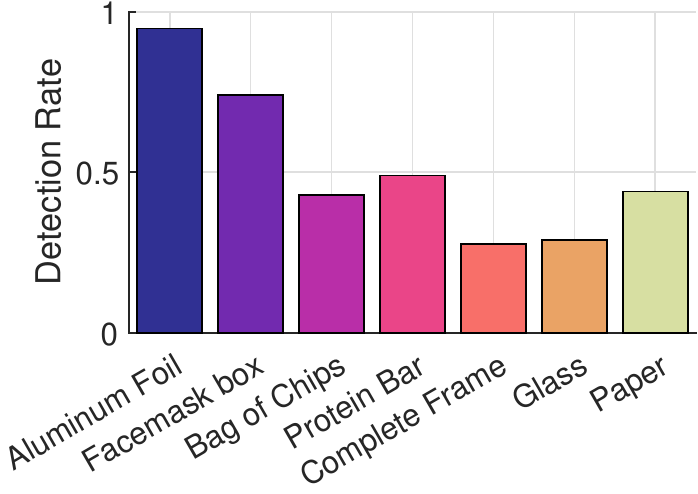}
 	\centering
 	\caption{Detection with different daily materials}
 	\label{fig:detection_passive_materials}
 \end{minipage}
 \hfill
 	\begin{minipage}{0.23\textwidth}
		\centering
		\includegraphics[width=1\textwidth]{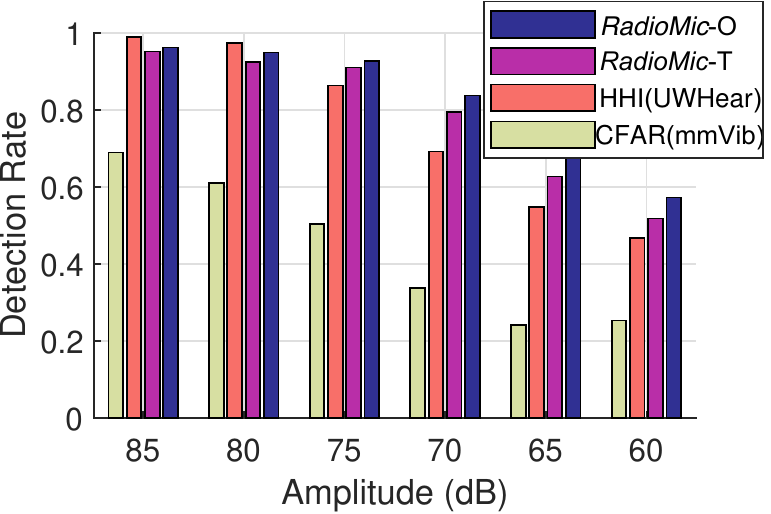}
		\caption{Detection at different sound levels}
		\label{fig:detection_passive_amplitude}
	\end{minipage}
 \end{figure*}

\head{\dlmodule Implementation:} 
To create synthetic data for training \dlmodule, we play a frequency sweep at various locations, and extract the radio channel response $h$. To account for fluctuations,
we apply a piecewise linear fit to the measured $h$, and added randomness to capture fluctuations.To capture noise
characteristics, we collect data in an empty room without any movement. Noise from each range bin is extracted, which are then added to
the synthetic data, with varying scaling levels to account for different locations. As the input dataset, we use
LibriSpeech \cite{panayotov2015librispeech} corpus.

Inputs to the neural network are taken as 128
samples ($\approx 1 s$), where only the middle 32 samples ($\approx 0.25s$)
are used for reconstruction and error calculation. 
We implement \dlmodule in PyTorch, and use an NVIDIA 2080S GPU
with CUDA toolkit for training. We use L2 loss between the real and estimated spectrograms.
Prior to calculating the error, we apply a mapping of $\log(1+G)$, where $G$
denotes a spectrogram. We randomly select 2.56 million spectrograms for
creating synthetic input/output data, and train the network for 10 epochs,
with Rmsprop algorithm.

\section{Experiments and Evaluation}
\label[]{sec:analysis}

We benchmark different modules of \sysname in multiple places, such as office space, home and an 
acoustically insulated chamber, and compare \sysname with the state-of-the-art approaches, followed by two case studies in \S\ref{sec:casestudies}.

\head{Evaluation Metrics:} 
It is non-trivial to evaluate the quality of sound. 
There are quite many different metrics in the speech processing literature, and we borrow some of them. Specifically, we adopt perceptual evaluation of speech
quality (PESQ) \cite{rix2002perceptual}, log-likelihood ratio (LLR)
\cite{quackenbush1988objective} and short-time intelligibility index (STOI)
\cite{taal2011shorttime}. PESQ tries to extract a quantitative metric that can
be mapped to mean-opinion-score, without the need for user study, and reports
values from 1 (worst) to 5 (best). LLR measures the \textit{distance} between two signals, and
estimates values in 0 (best) to 2 (worst). STOI is another measure of \textit{intelligibility}
of the sound, and reports values in $(0,1)$ with 1 the best. 
We avoid reporting SNR between the reference and reconstructed sound, as it does not correlate with human perception
\cite{quackenbush1988objective}. 
Rather, we report SNR using the noise energy estimated from silent moments, as it is used by \sysname during raw sound data extraction and it gives a relative idea on the amount of noise suppression.
Furthermore, we also visualize spectrograms and provide
sample files\footnote{Sample sound files recovered by \sysname can be found here: \url{https://bit.ly/radiomicsamples}.} to better illustrate the outputs of \sysname.

\subsection{Sound Detection Performance}
\label[]{subsec:evaldetectionlocalization}

\head{Overall Detection:}
Our data collection for detection analysis includes random motions, such as
standing up and sitting repeatedly, walking, running, and rotating in place, as 
well as static reflectors in the environment, and human bodies in front of the 
radar. On the other hand, we also collect data using multiple sound and music 
files with active and passive sources. 
More importantly, we have also collected motion and sound data at the same time
to see if \sysname can reject these interferences successfully.

To illustrate the gains coming from the proposed \soundmetric, we implement and
compare with existing methods: 
1) HHI (UWHear \cite{wang2020uwhear}): UWHear uses HHI, which requires some training to select an
appropriate threshold. 
2) CFAR (mmVib \cite{jiang2020mmvib}): The method used in \cite{jiang2020mmvib} requires
knowing the number of vibration sources a prior, and extracts the highest
peaks. To imitate this approach and provide a reasonable comparison, we apply CFAR detection rule at various threshold levels, and remove the detections around DC to have a
fairer comparison. 
Additionally, we
also compare hard thresholding (\sysname-T) with the outlier-based detector (\sysname-O). 
Note that the detector in \cite{xu2019waveear} 
requires extensive training, and is not applied here. 
We provide the receiver-operating characteristics (ROC) curve for all methods in Fig. \ref{fig:detectorroc}. 
As can be seen, while \sysname-T is slightly worse than \sysname-O, the other methods fail to
distinguish random motion from the vibration robustly, which prevents them from practical applications as there would be arbitrary motion in
the environment. 

\begin{figure}[t]
\centering

\subfloat[LLR]{
	\centering
	\includegraphics[width=0.21\textwidth]{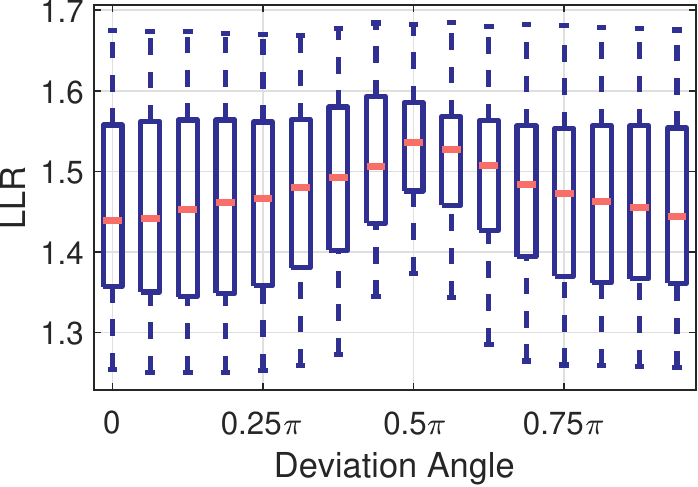}
	\label{fig:lineprojection_llr}
}
\hfill
\subfloat[STOI]{
	\centering
	\includegraphics[width=0.21\textwidth]{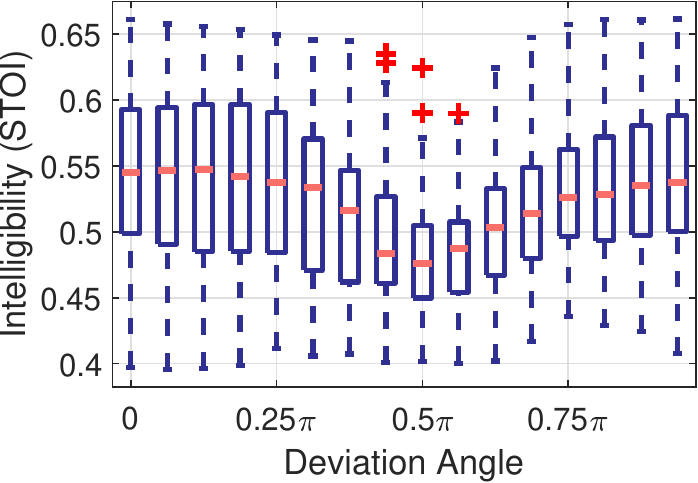}
	\label{fig:lineprojection_stoi}
}
\caption{Verification of line projection}
\label[]{fig:verification_lineprojection}
	\end{figure}

\head{Operational Range:} 
We investigate the detection performance of \sysname at different distances
azimuth angles (with respect to the radar) using an active source (\textit{a pair of speakers}) and a passive
source (\textit{aluminum foil of size $4\times 6$} inches). We use 5 different
sound files for each location, three of which are music files, and
two are human speech. 
As shown in Fig. \ref{fig:detection_heatmap_overall}, \sysname can robustly detect sound up to 4m
in an active case with 91\% mean accuracy, and up to $2m$ in the passive source case with \%70 accuracy, both with a field of view of 90$^\circ$. Passive source
performance is expected to be lower, as the vibration is
much weaker. 

\begin{figure*}[t]
\centering
\subfloat[SNR]{
	\centering
	\includegraphics[width=0.22\textwidth]{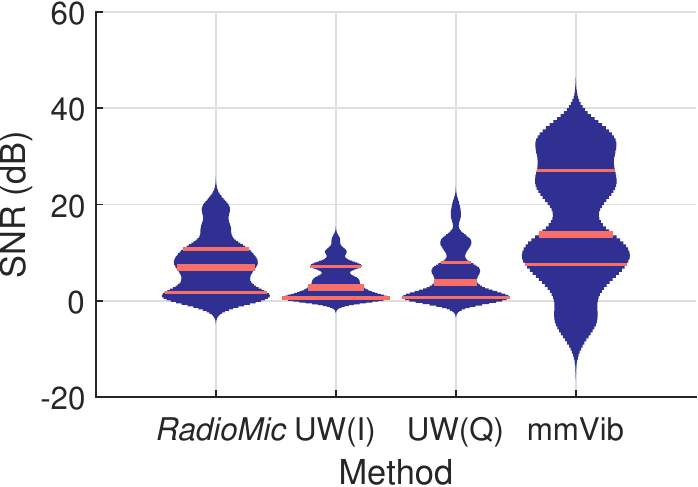}
	\label{fig:methodcomparison_snr}
}
\hfill
\subfloat[PESQ]{
	\centering
	\includegraphics[width=0.21\textwidth]{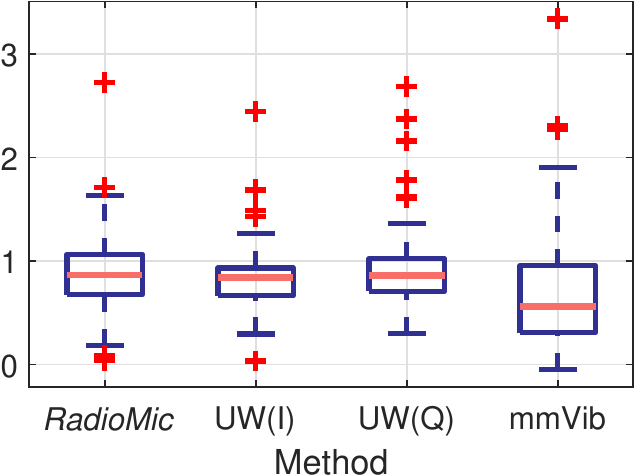}
	\label{fig:methodcomparison_pesq}
}
\hfill
\subfloat[LLR]{
	\centering
	\includegraphics[width=0.22\textwidth]{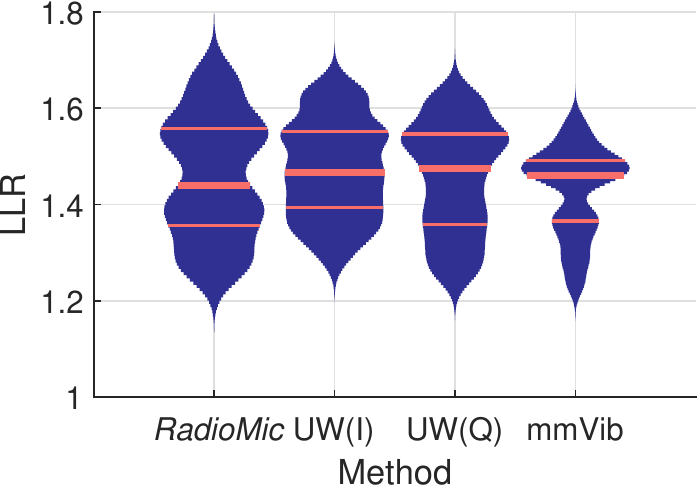}
	\label{fig:methodcomparison_llr}
}
\hfill
\subfloat[STOI]{
	\centering
	\includegraphics[width=0.22\textwidth]{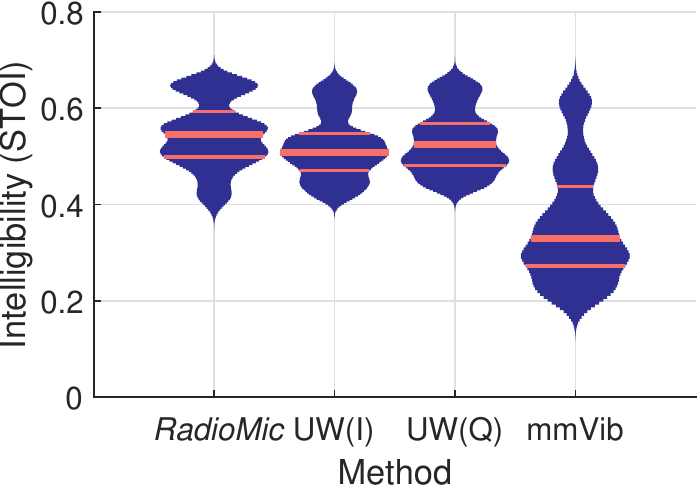}
	\label{fig:methodcomparison_stoi}
}
\caption{Comparison with UWHear \cite{wang2020uwhear} and mmVib \cite{jiang2020mmvib}. \rm{UW(I) and
UW(Q) denotes in-phase and quadrature signals extracted by UWHear
\cite{wang2020uwhear} respectively. RANet and diversity combining in
\sysname is not applied for this comparison. Horizontal lines on violin plots represent 25th, 50th and 75th percentile, respectively. Box plot is used for (b) due to outliers.}}
\label[]{fig:method_comparison_main}
\end{figure*}

\begin{figure*}[t]
\centering
\subfloat[SNR]{
	\centering
	\includegraphics[width=0.22\textwidth]{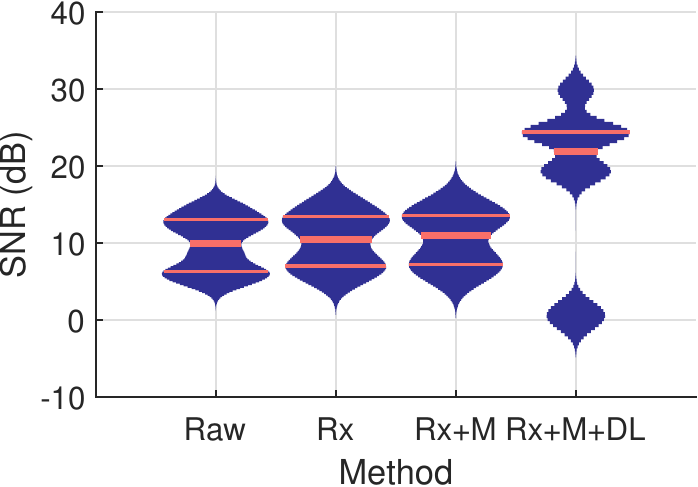}
	\label{fig:overall_snr}
}
\hfill
\subfloat[PESQ]{
	\centering
	\includegraphics[width=0.22\textwidth]{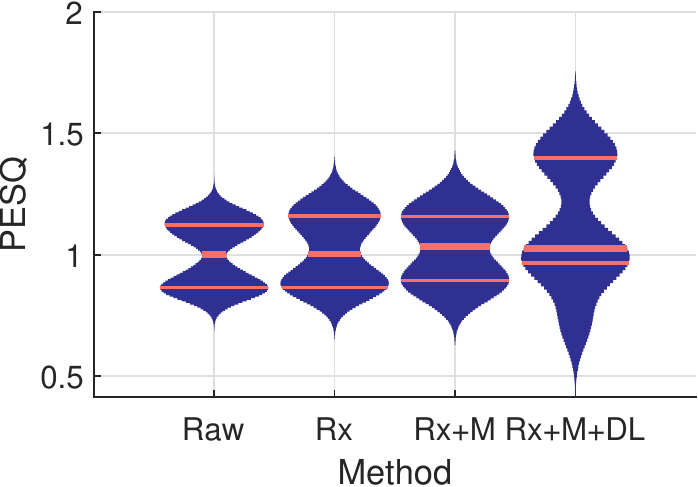}
	\label{fig:overall_pesq}
}
\hfill
\subfloat[LLR]{
	\centering
	\includegraphics[width=0.22\textwidth]{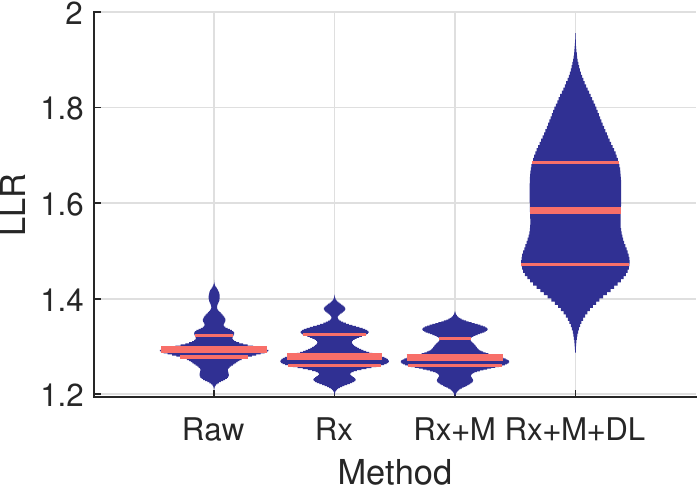}
	\label{fig:overall_llr}
}
\hfill
\subfloat[STOI]{
	\centering
	\includegraphics[width=0.22\textwidth]{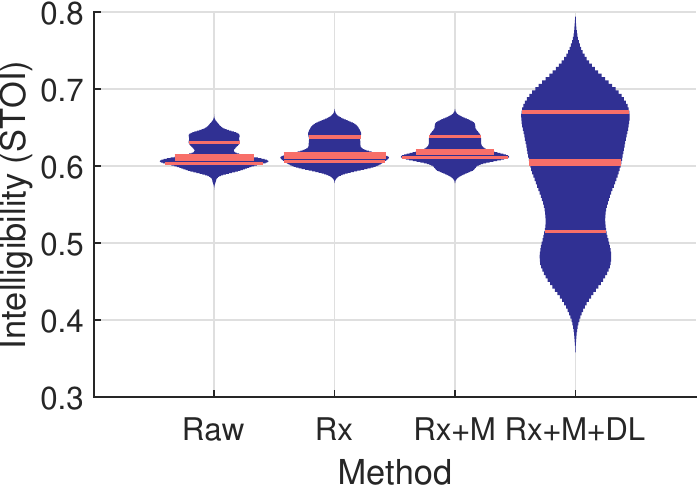}
	\label{fig:overall_stoi}
}
\caption{Overall performance of \sysname with gains from multiple components. \rm{Rx: receiver combining; Rx+M: receiver and multipath combining; Rx+M+DL: the end results.}}
\label[]{fig:overall_performance}
\end{figure*}

\head{Materials of Passive Sources:} 
To further evaluate detection performance of \sysname with passive materials, we
conduct experiments with additional daily materials, such as picture frames,
paper bags, or bag of chips. As shown in Fig.
\ref{fig:detection_passive_materials}, many different materials enable sound detection
using \sysname. 
And even at a lower rate, some sound signal is detected for
particular instances as the evaluation is done with frames with $40 ms$ duration. The performance could be improved by temporal
smoothing, if the application scenario requires a lower miss-detection ratio. 

\head{Sound Levels:} We also investigate the effect of sound amplitude on the detection ratio from a passive object. 
We reduce the amplitude of the test files from 85 dB to 60 dB, with a 5 dB step, and measure the detection rates in Fig. \ref{fig:detection_passive_amplitude}. As seen, \sysname outperforms existing approaches, especially when the amplitude is lower. 
Detecting sound at even lower levels is a common challenge for non-microphone sensing due to the extremely weak vibrations \cite{davis2014visual,wang2020uwhear,xu2019waveear}.

\subsection{Sound Reconstruction Performance}
\label[]{subsec:eval_reconstruction}

\head{Raw Reconstruction:} 

We first experimentally validate the mathematical model provided
in \S\ref{sec:theory}. 
Our main assertion is that, the sound signal could
be approximated with a linear projection on IQ plane, where the optimum angle
could be obtained by signal-to-noise maximizing projection. 
We test multiple projection angles deviated from the \sysname's
optimum angle, and generate results using LLR and STOI for
these angles. 
Our results in Fig. \ref{fig:verification_lineprojection} indicate the best results for projecting at $0^{\circ}$, and the
worst at $90^{\circ}$ with a monotonic decrease in between, which is consistent with the line model. This further indicates that an optimal scheme can achieve higher
performance than an arbitrary axis, as done in \cite{wang2020uwhear}.

\head{Comparative Study:} 
To compare our method with existing works, we employ only the raw reconstruction without RANet. 
The results are portrayed with various metrics in Fig. \ref{fig:method_comparison_main}. We provide
results with respect to in-phase (I) and quadrature (Q) part of UWHear
\cite{wang2020uwhear} as they do not provide a method to combine/select between
the two signals. For this comparison, we use an active speaker at 1m away,
with various azimuth angles.
Overall, \sysname outperforms both of the methods, for just the raw sound reconstruction, and it further improves the quality of the sound with additional processing blocks (Fig. \ref{fig:overall_performance}). 
The average SNR of mmVib \cite{jiang2020mmvib} is slightly higher. This is because SNR is calculated without a
reference signal, and drifts in the IQ plane boosts the SNR value of mmVib, as the later samples seem to have nonzero mean. However, the more important metrics, intelligibility score and
LLR, are much worse for mmVib, as it is not
designed to monitor the vibration over a long time, but for short frames.

\head{Overall Performance: }
With gains from diversity combining and deep learning, we provide the overall performance of \sysname in Fig.
\ref{fig:overall_performance}. We investigate the effect of each component on a dataset using a \textit{passive source}. Overall, each of the additional diversity
combining schemes improves the performance with respect to all metrics. At the
same time, \dlmodule reduces the total noise levels significantly (Fig. \ref{fig:overall_snr}) and increases PESQ (Fig. \ref{fig:overall_pesq}). 
However, as in Fig. \ref{fig:overall_llr}, \dlmodule yields a
worse value with LLR, which is due to the channel inversion operation of $h$ applied on
the radar signal. While an optimal channel recovery operation is demanded, \dlmodule is trained on multiple channel responses and only
approximates to $h$. Consequently, the channel inversion applied by \dlmodule is expected to be sub-optimal. Lastly, STOI metric (Fig. \ref{fig:overall_stoi}) shows a higher variation, which is
due to high levels of noise in the sample audio files in the input. 
In case of large noise, we have observed that \dlmodule learns to combat the effect
of noise $w$, instead of inverting $h$, and outputs mostly empty signals, which
could also be observed by the distribution around 0 dB in Fig. \ref{fig:overall_snr}. 
While when there is enough signal content, \dlmodule improves the
intelligibility further.

\begin{figure}
\centering
\subfloat[Locations]{
	\includegraphics[width=0.215\textwidth]{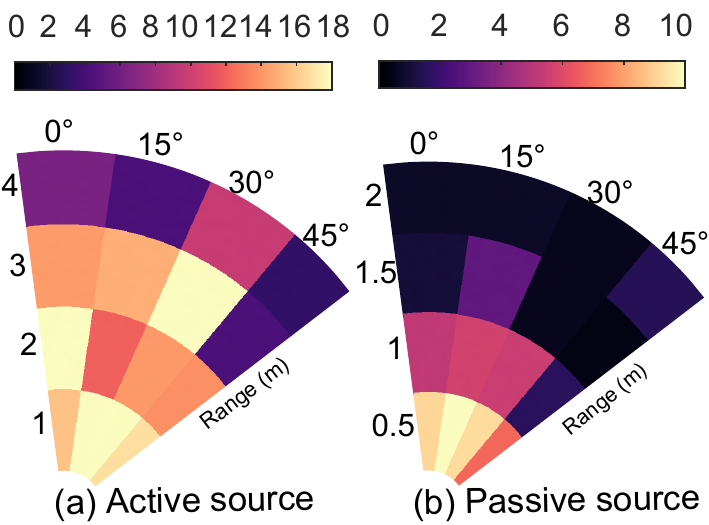}
	\label{fig:snr_heatmap_overall}
}
\hfill
\subfloat[Amplitudes]{
	\centering
	\includegraphics[width=0.215\textwidth]{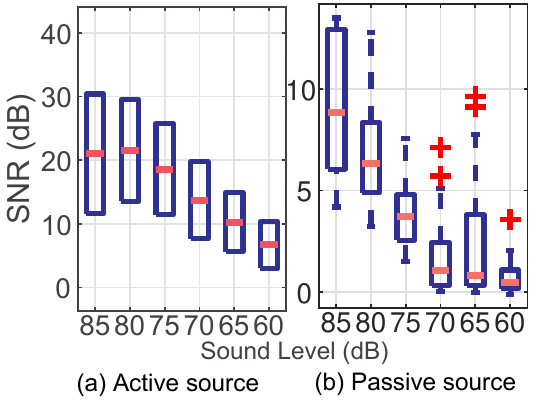}
	\label{fig:distance_amplitude_snr}
}
\caption{Recovered sound SNR (a) at different locations and (b) with different sound amplitudes.}
\label[]{fig:snr_heatmap_amplitude}
\end{figure}

\head{Distances and Source Amplitudes:} To investigate sound recovery from varying locations
and angles, we provide two heatmaps in Fig. \ref{fig:snr_heatmap_overall} to show the raw SNR output for active and passive sources. 
Similar to sound detection in Fig. \ref{fig:detection_heatmap_overall}, nearby locations have
higher SNR, allowing better sound recovery, and the dependency with
respect to the angle is rather weak. 
Increasing distance
reduces the vibration SNR strongly, (\eg, from 20 dB at 1m to 14 dB at 2m for an active
source) possibly due to the large beamwidth of our
radar device and high
propagation loss. 

We then test both active and passive sources at various sound levels. In Fig.
\ref{fig:distance_amplitude_snr}, we depict the SNR with
respect to sound amplitude, where the calibration is done by measuring the sound
amplitude at 0.5m away from the speakers, at $300$Hz.  
Generally, the SNR decreases with respect to decreasing sound levels. 
And at similar sound levels, a passive source, aluminum foil, can lose up to 10
dB compared to an active source. 
{In addition, \sysname retains a better
SNR with decreasing sound levels than increasing distance (Fig. \ref{fig:snr_heatmap_overall}), which indicates that the limiting factor for large distances is not the propagation loss, but the reflection loss, due to relatively smaller surface areas.} 
Hence, with more directional beams (\eg~transmit beamforming, or directional antennas), effective range of the \sysname could be improved, as low sound amplitudes also look promising for some recovery.

\begin{figure}
 	\includegraphics[width=0.45\textwidth]{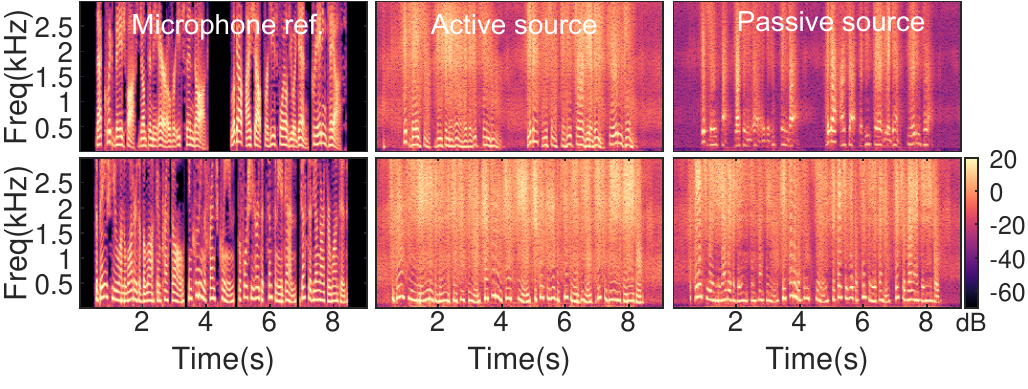}
 	\centering
 	\caption{Spectrogram comparison of \sysname outputs and a microphone. \rm{Two rows correspond to the synthesized speech of two different sentences. Passive source is a small aluminum foil, whereas active is a loudspeaker.}}
 	\label{fig:main_spectrograms_comparison}
 \end{figure}

\head{Active vs. Passive Comparison:} In order to show potential differences
 between the nature of active and passive sources, and present the results in a more perceivable way, we provide six spectrograms 
 in Fig. \ref{fig:main_spectrograms_comparison}, which are extracted by using two different synthesized audio files. In this setting, the passive source 
 (aluminum foil) is placed at 0.5m away and active source is located at 1m. As shown, 
 active source (speaker diaphragm) have more content in the lower frequency 
 bands, whereas passive sound results in more high frequency content, due to 
 the aggressive channel compensation operation on $h$. 
 More detailed comparisons are provided in Table 
 \ref{tab:active_passive_comparison}.

\begin{figure}[t]
\centering
\includegraphics[width=0.35\textwidth]{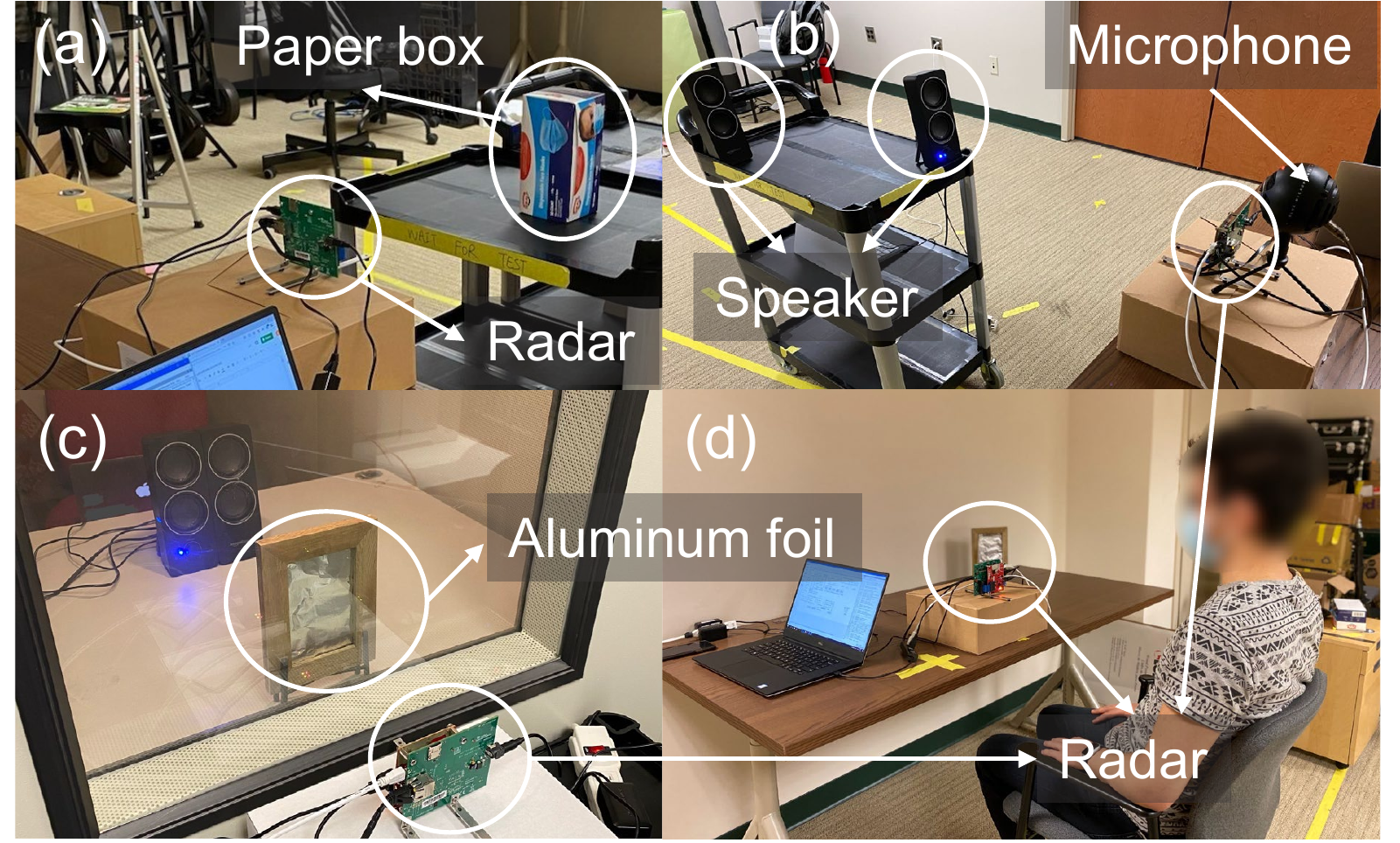}
\caption{Example setups. \rm{(a) Passive source; (b) Multiple speakers; (c) Insulated chamber; (d) Sensing from throat.}}
\label[]{fig:exp-setups}
\end{figure}

\begin{figure*}[t]
 	\centering
 \begin{minipage}{0.32\textwidth}
	\includegraphics[width=1.0\textwidth]{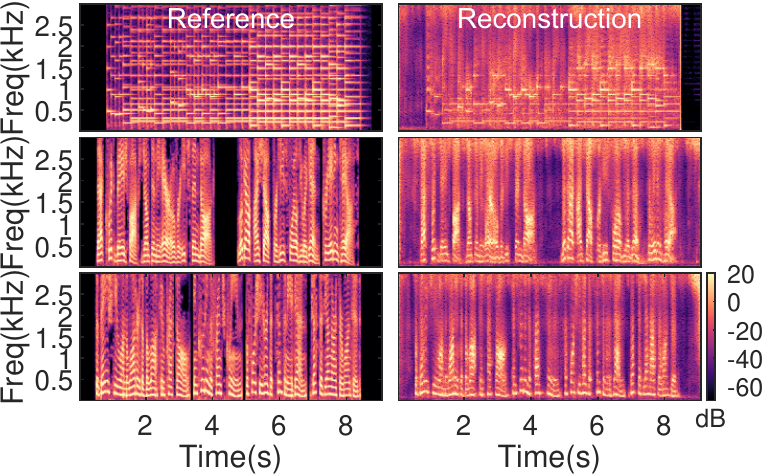}
	\caption{Through-wall spectrograms. \rm{Left: microphone reference; Right:
	reconstructed results. Top row also includes a music file.}}
	\label[]{fig:case-nlos}
 \end{minipage}
 \hfill
 \begin{minipage}{0.32\textwidth}
	\includegraphics[width=1.0\textwidth]{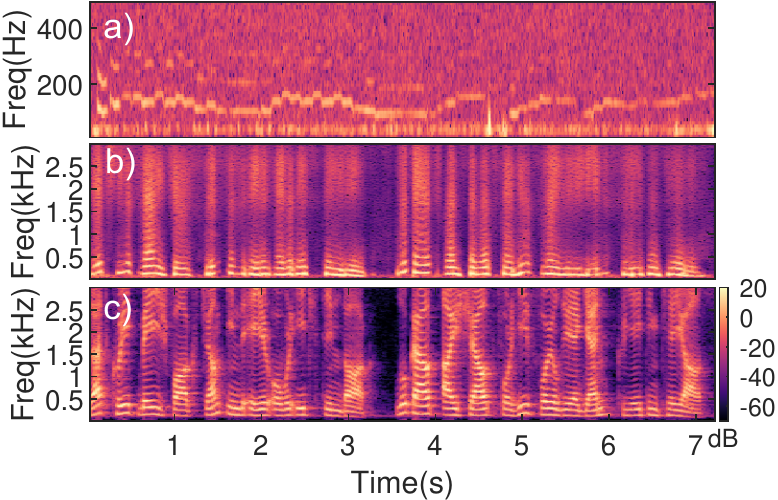}
	\caption{Recovery from throat. \rm{\sysname spectrogram of a) humming a song around 60 dB, and b) speaking, c) Microphone spectrogram for case b). }}
	\label[]{fig:throat_main}
 \end{minipage}
 \hfill
 	\begin{minipage}{0.32\textwidth}
		\centering
		\includegraphics[width=1\textwidth]{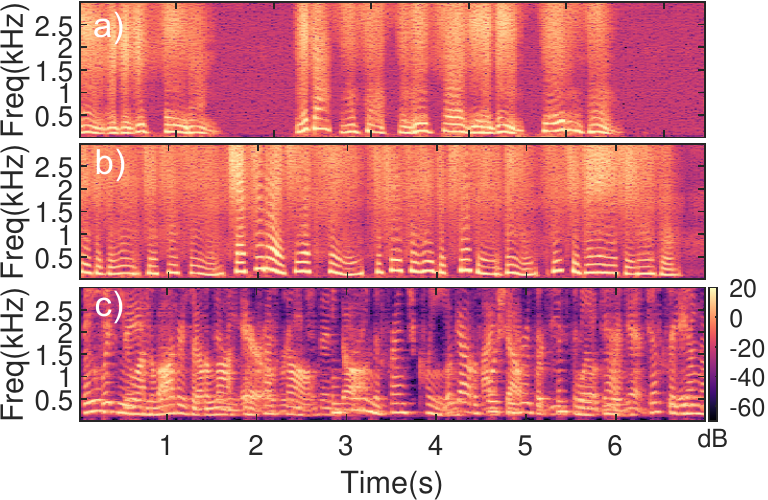}
	 	\caption{Multiple source separation. \rm{Spectrograms of \sysname for a) source \#1 and b) source \#2, c) Microphone spectrogram with mixed sound.}}
	 	\label{fig:multiplescource}
	\end{minipage}
 \end{figure*}

\head{LOS vs. NLOS Comparison:}
We further validate \sysname in NLOS operations. 
To that end, in addition to our office area, we conduct
experiments in an insulated chamber (Fig. \ref{fig:exp-setups}c), which has a double glass layer on its side.
This scenario is representative of expanding the range of an
IoT system to outside rooms from a quiet environment. 
In this particular scenario, we test both the passive
source (\eg~aluminum foil), and the active source (\eg~speaker). 
As additional layers would attenuate the RF reflection signals further, we test NLOS setup at slightly closer distances, with active speaker at 80 cm and the passive source at 35 cm away. 
Detailed results are in Table \ref{tab:active_passive_comparison}, with visual results in Fig. \ref{fig:case-nlos}. As seen, insulation layers does not affect \sysname much, and LOS and NLOS settings perform quite similar. Some metrics even show improvement in NLOS case due to shorter distances.

\begin{table}
	\caption{Active vs. Passive Source Comparison}
	\label[]{tab:active_passive_comparison}
	\centering
	\begin{tabular}{ l || c | c | c | c}
		\hline
		Setup & SNR & PESQ & LLR & STOI \\
		\hline
		LOS, Active & 24.7 & 0.84 & 1.61 & 0.55 \\
		LOS, Passive & 10.4 & 1.20 & 1.57 & 0.61 \\
		NLOS, Active & 29.4 & 1.12 & 1.52 & 0.58 \\
		NLOS, Passive & 8.8 & 1.36 & 1.57 & 0.64\\
		\hline
	\end{tabular}
\end{table}

\head{Sound Recovery from Human Throat:} Lastly, we show how \sysname can also capture vocal
folds vibration from human throat, as another active source. 
We start with humming in front of the speaker at a quiet 60 dB level, and show the results in \ref{fig:throat_main}a. After this observation, we collect multiple recordings from a user, where the setup is given in Fig. \ref{fig:exp-setups}d. In Fig.
\ref{fig:throat_main}b and \ref{fig:throat_main}c, we provide the \sysname and microphone spectrograms. Although \sysname is not trained with a frequency
response $h$ from human throat, it can still capture some useful signal content. On the other hand, we noticed that intelligibility of such speech is rather low, comparing to other sources, and \dlmodule sometimes does not estimate the actual speech. 
Prior work \cite{xu2019waveear} focuses on extracting sound from human throat,
and with extensive RF data collection, they have shown feasibility of better
sound recovery from throat. We believe the performance \sysname regarding human throat could be improved as well with a massive RF data from human throat. 
We leave these improvements for future and switch our focus to another application of sound liveness detection of human subjects in the next section. 

\section{Case Studies}
\label[]{sec:casestudies}

In this section, we first show \sysname for multiple source separation, and then extend it to classify sound sources.

\subsection{Multiple Source Separation}
\label[]{subsec:case-multiplesource}
 
Separation of multiple sound sources would enable multi-person sensing, or improved robustness against interfering noise sources. 
In order to illustrate feasibility, we play two different speech files simultaneously from left and right channels of the stereo speakers. As in Fig. \ref{fig:exp-setups}b, we place right speaker at 0.75m, and left speaker at 1.25m. We  
provide the results in Fig. \ref{fig:multiplescource}, which include two
spectrograms extracted by \sysname, along with a microphone spectrogram. 
As seen, microphone spectrogram extracts mixture of multiple sources,
and is prone to significant interference. In contrast, \sysname signals
show much higher fidelity, and two person's speech can be separated from each
other well. 
Previous work UWHear \cite{wang2020uwhear} specifically focuses on the problem of sound separation and demonstrates good performance using UWB radar. 
\sysname excels in achieving more features in one system, in addition to the source separation capability. 
And we believe there is a
great potential to pursue higher fidelity by using
\sysname \textit{in tandem with a single microphone}, which we leave for future investigation.

\begin{figure}
	\includegraphics[width=0.45\textwidth]{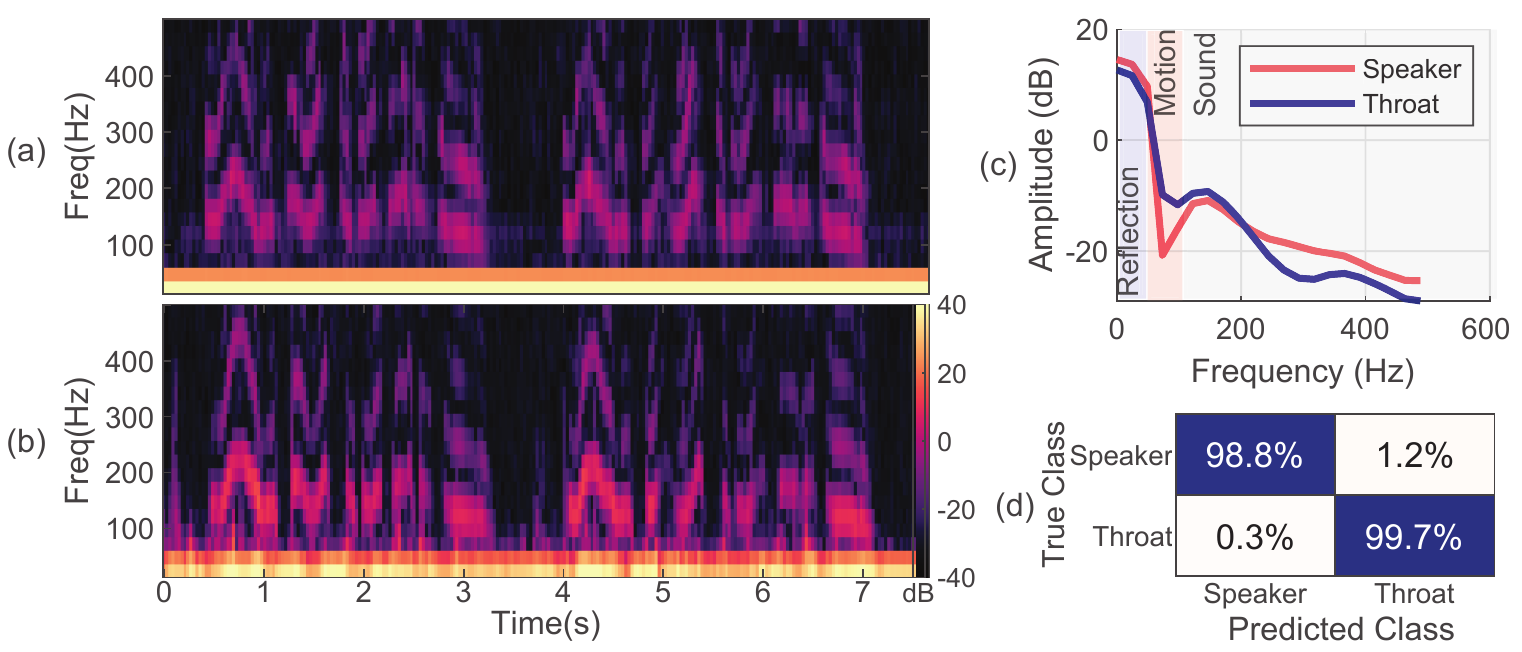}
	\centering
	\caption{(a) Speaker spectrogram, (b) throat spectrogram, (c) Power delay profile extracted from (a,b), (d) confusion matrix for classification.}
	\label[]{fig:spectrumcomparisonradarmic}
\end{figure}

\subsection{Sound Liveliness Detection}
\label[]{subsec:case-soundgeneration}
As another application, we investigate feasibility of sound source
classification. 
As \sysname senses at 
\textit{source} 
of the sound, it captures the additional physical characteristics of the 
\textit{sound generation mechanism} simultaneously. 
Starting with this observation, we investigate the question: \textit{Is it possible
to differentiate the source of a sound between a human and an inanimate source like a speaker?} 
This is a critical application as it is well-known that today's microphones all suffer from inaudible attacks \cite{zhang2017dolphinattack} and replay attacks \cite{roy2017backdoor} due to the hardware defects. 

Our results show that \sysname can enable sound liveliness detection, with unprecedented response times. 
In our experiment, we ask a user to recite five different
sentences, in two different languages, and we record the speech using a
condenser microphone. 
Afterwards, we play the same sound through speakers at the same
distance, at a similar sound level, and capture \sysname output.

We first provide the comparison of two spectrograms in
Fig.\ref{fig:spectrumcomparisonradarmic}(a,b), and the average over time in Fig. \ref{fig:spectrumcomparisonradarmic}(c). 
From the figures, we make three observations: 
1) As illustrated by \textit{reflection} band in Fig. \ref{fig:spectrumcomparisonradarmic}(c), the human throat shows a weaker amplitude around DC component. 
This is because the reflection coefficients of speakers and human throat vary significantly, a phenomenon utilized for material sensing \cite{wu2020msense}. 
2) Due to minute body motions and the
movement of the vocal tract, the reflection energy of human throat varies more over time and
has stronger sidelobes, which could be seen in the frequency band labeled as \textit{motion} in Fig. \ref{fig:spectrumcomparisonradarmic}(c).
3) Due to skin layer
between vocal cords and the radar, human throat applies stronger low-pass filtering on the vibration compared to speakers, as labeled as \textit{sound} in Fig. \ref{fig:spectrumcomparisonradarmic}c, which relates to the frequencies of interest for sound.

Then to enable \sysname for liveness detection, we implement a basic
classifier based on these observations.
We propose to use the ratio of the energy in motion affected bands (35-60 Hz)
over the entire radar spectrogram as an indicator for liveness. 
As shown in Fig. \ref{fig:spectrumcomparisonradarmic}(d), \sysname can classify the sources with $95\%$ accuracy with
only \textbf{40 ms} of data, which increases to $99.2\%$ by
increasing to $320ms$. 
We believe \sysname promises a valuable application here as it can sense the sound and classify the source \textit{at the same time}, and we plan to investigate it thoroughly as next step. 

\section{Related Work}
Wireless sensing has been an
emerging field \cite{kotaru2015spotfi,wang2015understanding,jiang2020towards,xie2020combating,chen2019widesee,zheng2019zero,wang2018promise, liu2019wireless,ma2019wifi}, with many applications including lip motion recognition and sound sensing. 
For example, voice recognition \cite{khanna2019through}, pitch extraction
\cite{chen2017detection}, and vocal cords vibration detection
\cite{hong2016time-varying} have been investigated using the Doppler radar
concept. 
Contact based RF sensors \cite{eid2009ultrawideband,birkholz2018non} have also been investigated, which
require user cooperation. Likewise, WiHear \cite{wang2014we} captures signatures of lip motions with WiFi, and matches different sounds, with a limited dictionary. 

A pioneer work \cite{wei2015acoustic} uses a 2.4 GHz SDR to
capture sound, while recently mmWave has been more widely employed. 
Some fundamentals have been established in \cite{li1996millimeter} between sound and mmWave. Recently, 
WaveEar \cite{xu2019waveear} uses
a custom built radar with 16 Tx and 16 Rx antennas to capture vocal folds
vibration, and reconstructs sound using a deep learning based approach. 
UWHear \cite{wang2020uwhear} focuses on 
separating multiple active sound sources with a UWB radar. mmVib \cite{jiang2020mmvib} does not directly recover sound but measures machinery vibration using a mmWave
radar. 
All of these approaches only focus
on an active vibration source.
Some introductory works study feasibility of sound sensing with passive
sources \cite{rong2019uwbradar,guerrero2020microwave}, but do not
build a comprehensive system. 
Moreover, existing works mainly focus on lower frequency signals, and do not address the fundamental limitations in
high frequency.

\head{Ultrasound-based:} 
Ultrasound based methods are used to capture lip motion \cite{jennings1995enhancing}, recognize the speaker \cite{kalgaonkar2008ultrasonic},
synthesize speech \cite{toth2010synthesizing}, or
enhance sound quality \cite{lee2019speech}. 
These approaches usually have very limited range, and require a prior dictionary, as
the motion cannot be related to sound immediately. 
We note that \sysname differs fundamentally from acoustic sensing \cite{zhang2017dolphinattack,wang2016device,mao2018aim,yun2015turning}, which leverages sound for sensing, rather than recovering the sound itself. 

\head{Light based:} 
VisualMic \cite{davis2014visual} recovers sound from the
\env (\eg~bag of chips) using high-speed cameras. 
The similar phenomenon is exploited using lamp \cite{nassi2020lamphone}, laser \cite{muscatell1984laser}, lidar \cite{sami2020spying}, and depth cameras \cite{galatas2012audio-visual}. 
Comparing to \sysname, these works usually need expensive specialized hardware. 

\head{IMU-based:} 
Inertial sensors are also used for sound reconstruction \cite{zhang2015accelword,roy2016listening,michalevsky2014gyrophone}. 
All these methods sense the sound at
\dst like contact microphones, and has similar drawbacks to microphones, in addition to their limited
bandwidth.

\section{Conclusion}
\label[]{sec:conclusion}
In this paper, we propose \sysname, a mmWave radar based sound sensing system,
which can reconstruct sound from sound sources and passive objects in the environment.
Using the tiny vibrations that occur on the object surfaces due to ambient
sound, \sysname can detect and recover sound as well as identify sound sources using a novel radio acoustics model and neural network. 
Extensive experiments in various settings show that \sysname outperforms existing approaches significantly and benefits many applications.

There are room for improvement and various future directions to explore. 
First, our system can capture the sound up to 4 meters, and higher
order frequencies start to disappear beyond 2 meters, mainly due to a relatively wide beamwidth of $30^{\circ}$ of our device. 
Other works focusing on extracting throat
vibration either use highly directional antennas (even up to $1
^{\circ}$) beamwidth \cite{chen2017detection}, very close distance (less than 40 cm) in \cite{khanna2019through}, 
many more antennas (\eg, 16$\times$16) \cite{xu2019waveear}.  
We believe that more advanced hardware and sophisticated beamforming could underpin better performance of \sysname.

Second, \dlmodule mitigates the fundamental limit of high-frequency deficiency, which currently uses 1-second window with limited RF training data. 
Better quality could be achieved by exploiting long-term temporal dependencies with a more complex model, given more available RF data. 
With more RF data from human throat, we also expect to see a better performance for human speech sensing. 

Third, we believe \sysname and microphones are complementary. Using suitable deep learning techniques, the side
information from \sysname could be used in tandem with microphone to achieve better performance of sound separation and noise mitigation than microphone alone. 

Lastly, with our initial results, it is promising to build a security system for sound liveness detection against side-channel attacks. 
Exploring \sysname with mmWave imaging and other sensing \cite{soli2021sleep} is also an exciting direction. 

\newpage
\balance

\bibliographystyle{ACM-Reference-Format}
\bibliography{RadioMic-arxiv}

\end{document}